\documentclass[prd,nofootinbib,twocolumn]{revtex4}

\usepackage{amssymb}
\usepackage{amsmath}
\usepackage{graphicx,subfigure}
\usepackage{longtable}
\usepackage{verbatim}
\usepackage{amsfonts}

\allowdisplaybreaks

\newcommand{\rep}[1]{\mathbf{#1}}%

\newcommand{\be}{\begin{equation}}
\newcommand{\ee}{\end{equation}}
\newcommand{\bea}{\begin{eqnarray}}
\newcommand{\eea}{\end{eqnarray}}
\newcommand{\beq}{\begin{equation}}
\newcommand{\eeq}{\end{equation}}
\def\beqa{\begin{eqnarray}}
\def\eeqa{\end{eqnarray}}
\newcommand{\bv}{\left(\begin{array}{c}}
\newcommand{\ev}{\end{array}\right)}
\newcommand{\no}{\nonumber}
\def\lsim{\mathrel{\rlap{\lower4pt\hbox{\hskip1pt$\sim$}}
    \raise1pt\hbox{$<$}}}         
\def\gsim{\mathrel{\rlap{\lower4pt\hbox{\hskip1pt$\sim$}}
    \raise1pt\hbox{$>$}}}         

\begin{document}

\begin{flushright}
DO-TH 12/22
\end{flushright}

\vspace*{-30mm}

\title{\boldmath 
$SU(3)$-Flavor Anatomy of Non-Leptonic  Charm Decays}

\author{Gudrun Hiller}
\email{gudrun.hiller@tu-dortmund.de}
\author{Martin Jung}
\email{martin2.jung@tu-dortmund.de}
\author{Stefan Schacht}
\email{stefan.schacht@tu-dortmund.de}
\affiliation{Institut f\"ur Physik, Technische Universit\"at Dortmund, D-44221 
Dortmund, Germany}

\vspace*{1cm}

\begin{abstract}
We perform a comprehensive $SU(3)$-flavor analysis of charmed mesons decaying to two pseudoscalar
$SU(3)$-octet mesons.
Taking into account $SU(3)$-breaking effects induced by the splitting of  the quark masses, $m_s \neq m_{u,d}$, 
we find that  existing data can be described by  $SU(3)$-breaking of the order $30\%$.
The requisite penguin enhancement to accommodate all data on CP violation 
tends to be even larger than the one extracted from $\Delta a_{CP}^{\rm dir}(K^+K^-,\pi^+\pi^-)$ alone, strengthening 
explanations beyond the standard model.
Despite the large number of matrix elements, 
correlations between CP asymmetries allow potentially to differentiate between different scenarios for the
underlying dynamics, as well as between the standard model and various extensions characterized by $SU(3)$ symmetry and its subgroups.
We investigate how improved measurements of
the  direct CP asymmetries  in 
singly-Cabibbo-suppressed decays can further substantiate the interpretation of the data.
We show that particularly informative are the asymmetries in
$D \to \pi^+ \pi^-$ versus
$D \to K^+ K^-$,  $D_s \to K_S \pi^+$ versus $D^+ \to K_S K^+$,
$D^+\to \pi^+ \pi^0$, $D\to\pi^0\pi^0$, and $D \to K_S K_S$.
\end{abstract}

\maketitle

\section{Introduction}
The recent measurements of 
direct CP violation in non-leptonic charm decays \cite{Aaij:2011in,Collaboration:2012qw,ICHEP12:Ko}
have been among the most exciting results in flavor physics in recent years.
The combined significance of CP violation in charm decays is $4.6\sigma$
\cite{Amhis:2012bh},
\beqa\label{eq:acp}
\Delta a_{CP}^{\rm dir}
&\equiv&
a_{CP}^{\rm dir}
(  D^0 \to K^+K^-)-a_{CP}^{\rm dir}
(D^0 \to \pi^+\pi^-)\no\\
&=&(-0.678\pm0.147) \cdot 10^{-2}\,,
\eeqa
where
\begin{align}\label{eq::acpdir}
a_{CP}^{\rm dir}(d) &= \frac{|{\cal{A}} (d)|^2 - |\bar {\cal{A}} (d)|^2}{
		|{\cal{A}} (d)|^2 + |\bar {\cal{A}} (d)|^2} 
\end{align}
for a decay $d$ of a $C=+1$ meson, where $\mathcal{A}(d)$ denotes the weak decay amplitude of the flavor eigenstates.
This measurement, together with the plethora of available charm data, 
makes an $SU(3)$-flavor symmetry analysis worthwhile,
aiming at an understanding within or beyond the standard model (SM).
The basics of such an analysis were laid out some time ago \cite{Kingsley:1975fe,Einhorn:1975fw,Altarelli:1974sc,Voloshin:1975yx,Quigg:1979ic}; however, the present situation allows for a much more complete analysis than was possible before.

The $SU(3)$ symmetry is known to be broken rather severely in charm decays, the most striking example given by $\Gamma(D^0\to K^+K^-)/\Gamma(D^0\to\pi^+\pi^-)\sim2.8$. However, as first pointed out in \cite{Savage:1991wu}, this does not necessarily imply
$SU(3)$ breaking on the amplitude level beyond the expected order of $\sim30\%$ (see also \cite{Chau:1991gx} for a similar analysis in the diagrammatic approach). Indeed, for a subset of decays it has been 
shown again recently that ``nominal'' $SU(3)$ breaking is sufficient to explain this ratio \cite{Pirtskhalava:2011va,Feldmann:2012js,Brod:2012ud}.

Here we address the following questions:
 \begin{itemize}
 \item[\it i]  How large is the requisite  $SU(3)$  breaking in charm decays?
 \item[\it ii]  How large is the requisite penguin, {\it i.e.,} triplet matrix element  enhancement to explain the observed CP violation?
 \item[\it iii]  Can we distinguish between new physics (NP) contributing to operators in different representations of $SU(3)$?  Which measurements would be particularly useful?
 \end{itemize} 

In our analysis we take into account
$SU(3)$-breaking  induced by the splitting in the quark masses $m_s \neq m_{u,d}$ to first order;
we compare our findings to
 the most complete, relevant data set of decays into two octet pseudoscalars and
 employ no further dynamical assumptions, 
 the combination of which is where we go  beyond previous works on SM CP violation,  
 {\it e.g.}, recently
 \cite{Brod:2011re,Pirtskhalava:2011va,Cheng:2012wr,Bhattacharya:2012ah,Feldmann:2012js,Li:2012cfa,Franco:2012ck,Brod:2012ud},
 besides a detailed assessment of the $SU(3)$-flavor anatomy of non-leptonic charm decays.

The plan of the paper is as follows:  The $SU(3)$-structure of hadronic two-body
charm decays  is given in Section~\ref{sec:su3-setup}. In Section~\ref{sec:mfv-fit} we present fits to the data assuming the SM. In particular, all CP violation is induced by the Cabibbo-Kobayashi-Maskawa (CKM) mixing matrix. This analysis hence holds
more generally in all models with this minimally flavor violating (MFV)-feature. 
 In Section~\ref{sec:pattern} we allow for CP violation from NP characterized by different $SU(3)$ representations. We identify patterns in observables that can guide towards
an identification of the underlying flavor dynamics. We conclude in Section~\ref{sec:end}.
In several appendices we give Clebsch-Gordan tables and subsidiary information.

\section{$\mathbf{SU(3)}$-decomposition 
\label{sec:su3-setup}}

We present the $SU(3)$-decomposition of various two-body charm decay amplitudes.
The requisite Clebsch-Gordan coefficients are obtained  using the tables and the program from
Refs.~\cite{de Swart:1963gc, Kaeding:1995vq,Kaeding:1995re}.

\subsection{$\mathbf{SU(3)}$-limit}

$SU(3)$-flavor symmetry allows to express the amplitudes of the various decays $d$ in
terms of reduced matrix elements $A_i^k$ as
\begin{align} \nonumber
{\cal{A}}_0(d) &= \Sigma\sum_{i,k} c_{d; ik} A_i^k\,,  & ~~~~(\mbox{SCS})\\ 
{\cal{A}}_0(d) &=V_{cs}^* V_{ud} \sum_{i,k} c_{d; ik} A_i^k\,,  & ~~~~(\mbox{CF}) \label{eq:su3amplitudes}\\
{\cal{A}}_0(d) &=V_{cd}^* V_{us} \sum_{i,k} c_{d; ik} A_i^k\,,   & ~~~~(\mbox{DCS})
\nonumber
\end{align}
where 
\beq
\Sigma   \equiv( V_{cs}^*V_{us}-V_{cd}^*V_{ud})/2\,.
\eeq

We employ the commonly used classification for non-leptonic two-body charm decays according to the
CKM hierarchy of their decay amplitudes:
Cabibbo-favored (CF) at order one, singly-Cabibbo-suppressed (SCS) at order
$\lambda$ and
doubly-Cabibbo-suppressed (DCS) at order $\lambda^2$ in the Wolfenstein expansion, where $\lambda\simeq 0.2$.

In Eq.~(\ref{eq:su3amplitudes})  $i,k$ label the representation of the final state and the $\bar u c \bar q q^\prime$  interaction Hamiltonian, respectively.
 The relevant tensor product of the latter is written as
\begin{equation}
\rep{3}\otimes\rep{\bar{3}}\otimes\rep{3}=\rep{3_1}\oplus \rep{3_2}\oplus
\rep{\bar{6}}\oplus\rep{15}\,.
\end{equation}
The initial $D$-meson ($D=(D^0,D^+,D_s)$) is an $SU(3)$ anti-triplet.
The  two pseudoscalar octets in the final state can be decomposed as
 $(\rep{8}\otimes\rep{8})_S=\rep{1}\oplus\rep{8}\oplus
\rep{27}$, where we symmetrized to account for Bose statistics and removed the dependence on the order of the final state mesons.
The resulting coefficients $c_{d;ij}$ are given in Table~\ref{tab::SU3limitoctet},
where we introduced $\tilde\Delta=(V_{cs}^*V_{us}+V_{cd}^*V_{ud})/(2\Sigma)\sim\lambda^4\sim 10^{-3}$, which characterizes 
the CKM suppression of direct CP violation in $D$ decays. Note furthermore that we combined the coefficients of the $\rep{3}_1$ and $\rep{3}_2$, as they have identical quantum numbers and therefore enter each amplitude with identical relative weight.
Our findings are in agreement with Ref.~\cite{Quigg:1979ic}; we disagree with
a recent calculation \cite{Pirtskhalava:2011va} in the sign of the $D^+\to K^+\pi^0$ 
amplitude\footnote{The authors of Ref.~\cite{Pirtskhalava:2011va} informed us that they agree with our expressions. The apparent sign differences are due to typos in their coefficient tables. \label{fn:sign}}.
The inclusion of decays into pseudoscalar singlets is
left for future work \cite{hjs13}.

\begin{table}
\begin{center}
\begin{tabular}{l|c|c|c|c|c}
\hline \hline
Decay $d$ & $A_{27}^{15}$ & $A_8^{15}$ &
$A_8^{\bar 6}$ & $A_1^3$ & $A_8^3$ \\\hline\hline
\multicolumn{6}{c}{SCS} \\\hline\hline
$D^0\rightarrow K^+ K^-$  & $\frac{3 \tilde\Delta +4  }{10 \sqrt{2}}$  & $\frac{\tilde\Delta -2  }{5 \sqrt{2}}$  & $\frac{1 }{\sqrt{5}}$  &  $\frac{\tilde\Delta }{2 \sqrt{2}}$  & $\frac{\tilde\Delta }{\sqrt{10}}$  \\\hline
$D^0\rightarrow \pi^+ \pi^- $ & $\frac{3 \tilde\Delta -4  }{10 \sqrt{2}}$   & $\frac{\tilde\Delta +2  }{5 \sqrt{2}}$   & $-\frac{1 }{\sqrt{5}}$    &  $\frac{\tilde\Delta }{2 \sqrt{2}}$    &  $\frac{\tilde\Delta }{\sqrt{10}}$   \\\hline
$D^0\rightarrow \bar{K}^0 K^0$ & $\frac{\tilde\Delta }{10 \sqrt{2}}$  &  $\frac{\sqrt{2} \tilde\Delta }{5}$   & $0$  & $-\frac{\tilde\Delta }{2 \sqrt{2}}$  & $\sqrt{\frac{2}{5}} \tilde\Delta$    \\\hline
$D^0 \rightarrow \pi^0 \pi^0$ & $\frac{7 \tilde\Delta -6 }{20}$  &  $-\frac{\tilde\Delta +2 }{10}$    & $\frac{1 }{\sqrt{10}}$    &  $-\frac{\tilde\Delta }{4} $  &  $-\frac{\tilde\Delta }{2 \sqrt{5}}$   \\\hline
$D^+ \rightarrow \pi^0 \pi^+$ & $\frac{\tilde\Delta -1 }{2}$  & $0$  & $0$  & $0$  & $0$  \\\hline
$D^+ \rightarrow \bar{K}^0 K^+$ & $\frac{\tilde\Delta +3  }{5 \sqrt{2}}$   &  $-\frac{\tilde\Delta -2  }{5 \sqrt{2}}$   & $ \frac{1 }{\sqrt{5}}$    & $0$  &  $\frac{3 \tilde\Delta }{\sqrt{10}} $  \\\hline
$D_s \rightarrow  K^0 \pi^+$ &  $\frac{\tilde\Delta -3  }{5 \sqrt{2}}$  & $-\frac{\tilde\Delta +2  }{5 \sqrt{2}}$   &  $-\frac{1 }{\sqrt{5}}$   & $0$  & $ \frac{3 \tilde\Delta }{\sqrt{10}}$  \\\hline
$D_s \rightarrow  K^+ \pi^0$ &  $\frac{2 \tilde\Delta -1}{5}$   & $ \frac{\tilde\Delta +2 }{10}$    & $ \frac{1 }{\sqrt{10}}$   & $0$  & $-\frac{3 \tilde\Delta }{2 \sqrt{5}}$  \\\hline\hline
\multicolumn{6}{c}{CF} \\\hline\hline
$D^0\rightarrow K^- \pi^+$ & $\frac{\sqrt{2}}{5}$ & $-\frac{\sqrt{2}}{5}$ & $\frac{1}{\sqrt{5}}$ & $0$ & $0$  \\\hline
$D^0\rightarrow \bar{K}^0 \pi^0$  & $\frac{3}{10}$    &  $\frac{1}{5}$   &  $-\frac{1}{\sqrt{10}}$   & $0$  & $0$  \\\hline
$D^+ \rightarrow \bar{K}^0 \pi^+$ &  $\frac{1}{\sqrt{2}}$   & $0$  & $0$  & $0$   & $0$   \\\hline
$D_s \rightarrow \bar{K}^0 K^+$ & $\frac{\sqrt{2}}{5}$  & $-\frac{\sqrt{2}}{5}$   & $-\frac{1}{\sqrt{5}} $  & $0$  & $0$  \\\hline\hline
\multicolumn{6}{c}{DCS} \\\hline\hline
$D^0 \rightarrow  K^+ \pi^-$ & $\frac{\sqrt{2}}{5}$  & $-\frac{\sqrt{2}}{5}$  & $\frac{1}{\sqrt{5}}$   & $0$  & $0$  \\\hline
$D^0 \rightarrow K^0 \pi^0$ & $\frac{3}{10}$  &  $\frac{1}{5}$   & $-\frac{1}{\sqrt{10}}$   & $0$  & $0$  \\\hline
$D^+ \rightarrow K^0\pi^+ $ & $\frac{\sqrt{2}}{5}$   & $-\frac{\sqrt{2}}{5}$    & $-\frac{1}{\sqrt{5}}$   & $0$  & $0$  \\\hline
$D^+ \rightarrow K^+ \pi^0$ &  $\frac{3}{10}$   &  $\frac{1}{5}$   & $\frac{1}{\sqrt{10}}$   & $0$  & $0$  \\\hline
$D_s \rightarrow K^0 K^+ $ &  $\frac{1}{\sqrt{2}}$   & $0$  & $0$  & $0$  & $0$ \\\hline \hline
\end{tabular}
\caption{The coefficients $c_{d;ij}$ of the decomposition into reduced matrix elements in the SU(3)-limit given in Eq.~(\ref{eq:su3amplitudes}). 
\label{tab::SU3limitoctet}}
\end{center}
\end{table}

\subsection{Breaking $\mathbf{SU(3)}$ \label{sec:su3break}}

Including $SU(3)$-breaking through  $m_s \neq m_{u,d}$, {\it i.e.}, leaving isospin symmetry intact, 
leads to $SU(3)$ breaking by a single representation $(\rep{8})$, see also \cite{Savage:1991wu,Hinchliffe:1995hz,Pirtskhalava:2011va}.
The effective Hamiltonian  contains at lowest order in the breaking the following
decompositions:
\begin{eqnarray} \nonumber
\rep{15}\otimes\rep{8}&=&\rep{42}\oplus\rep{24}\oplus\rep{15_1}\oplus\rep{15_2}
\oplus\rep{15'}\oplus\rep{\bar{6}}\oplus\rep{3}\,, \\
\rep{\bar{6}}\otimes\rep{8} &=& \rep{24}\oplus\rep{15}\oplus\rep{\bar{6}}\oplus
\rep{3}\,,\quad\\
\rep{3}\otimes\rep{8}&=&\rep{15}\oplus\rep{\bar{6}}\oplus\rep{3}\,.\nonumber
\label{eq:heffsu3break}
\end{eqnarray}
We consider   the CKM-leading terms in the flavor-breaking only\footnote{One should revisit this assumption once CP data become more precise.},
hence the $\rep{3}\otimes\rep{8}$ does not contribute. A notable exception is the decay $D^+\to\pi^0\pi^+$, where the omission of $\tilde \Delta$ would induce a spurious CP asymmetry. However, in the absence of new isospin violating interactions, the latter vanishes, see, {\it e.g.}, \cite{Buccella:1992sg,Grossman:2012eb}.
Furthermore, there are no non-zero matrix elements with the $\rep{15'}$.
The decay amplitudes  can then be written as
\begin{align} \label{eq:expansion}
{\cal{A}}(d)={\cal{A}}_0(d)+{\cal{A}}_{\rm X}(d) \, ,
\end{align}
where  the ${\cal{A}}_{\rm X}$ denote the flavor-breaking contributions.
We express them in terms of reduced matrix elements $B_i^j$,
\begin{align} 
{\cal{A}}_{\rm X} (d) &=\Sigma \sum_{i,j} c_{d; ij} B_i^j\,,  & ~~~~(\mbox{SCS})\label{eq:su3breakamplitudes}\\ 
{\cal{A}}_{\rm X}(d) &=V_{cs}^* V_{ud} \sum_{i,j} c_{d; ij} B_i^j\,,  & ~~~~(\mbox{CF}) \\
{\cal{A}}_{\rm X} (d) &=V_{cd}^* V_{us} \sum_{i,j} c_{d; ij} B_i^j\,.  & ~~~~(\mbox{DCS})\label{eq:su3breakamplitudes3}
\end{align}
analogous to Eq.~(\ref{eq:su3amplitudes}).

Our findings  for the coefficients $c_{d;ij}$  are given  in Table~\ref{tab:SU3breakoctetpre}.
We confirm the results given in Ref.~\cite{Pirtskhalava:2011va} except for  the sign
of the  $\Sigma $-terms in the $SU(3)$-breaking part of the $D \to \pi^0 \pi^0$ amplitude and
the overall sign of the $D^+\to K^+\pi^0$ amplitude\footnote{See footnote \ref{fn:sign}.}.

For $\tilde \Delta\to0$, the flavor structure of Eq.~(\ref{eq:heffsu3break}) leads 
to 15  $SU(3)$-breaking matrix elements in addition to the three from the $SU(3)$-limit. 
However, the coefficient matrix of Clebsch-Gordan coefficients
does not have full rank, implying  that not all matrix elements are physical. 
Since in fact the coefficient matrix has rank 11, we  go on and reduce the number of matrix elements by 7.
Specifically, the two $\rep{3}$ representations have again the same quantum numbers, allowing us to combine
the corresponding coefficients.
Using Gaussian elimination we can further remove 
$B_8^{\bar{6}_2}$, $B_8^{15_3}$, $B_{27}^{15_3}$, $B_{27}^{24_2}$, and $B_{27}^{42}$. 
We do not mix leading $SU(3)$ elements in the process, and keep their normalizations when subleading matrix
elements are absorbed. 
The $\tilde\Delta$-suppressed 
$SU(3)$-breaking matrix elements that appear in the course of the redefinitions are again neglected as they are  of
higher order in the power counting employed in this work.
Due to the reparametrizations, the sub- and superscripts on the 
$B_i^j$ cease to correspond to
the $SU(3)$-representations of the interaction and the final states. 
The appearing combinations have been normalized according to $\sqrt{\sum_i|c_i|^2}B^{\rm phys} = \sum_i c_i B_i^{SU(3)}$, 
in order not to change the relative normalization of the matrix elements.
The resulting coefficients $c_{d;ij}$  of the physical decomposition are given in Table~\ref{tab:SU3breakoctet}.
As anticipated, we end up with in total 13 independent matrix elements.
In the following analyses we use Eqs.~(\ref{eq:su3breakamplitudes})-(\ref{eq:su3breakamplitudes3})  with the coefficients from Table~\ref{tab:SU3breakoctet}.

\subsection{$\mathbf{SU(3)}$-breaking resistent sum rules} 

For $\tilde\Delta=0$ the SCS 8$\times$11 submatrix of Clebsch-Gordan coefficients including $SU(3)$ breaking in the physical parametrization has rank 7. Therefore, we know a priori that there is only one linear sum rule among the SCS amplitudes that remains valid after $SU(3)$ breaking. This sum rule is the well-known isospin relation \cite{Kwong:1993ri}
\footnote{The apparent discrepancy between Eq.~(\ref{eq:isospin}) and previous 
works~\cite{Quigg:1979ic,Kwong:1993ri}  is due to the
different conventions used, specifically
$\mathcal{A}\left(D^0\rightarrow \pi^0 \pi^0\right)_{here} =-\sqrt{2}
\mathcal{A}\left(D^0\rightarrow\pi^0 \pi^0\right)_{previous}$ and
$\mathcal{A}\left(D^+\rightarrow \pi^+ \pi^0\right)_{here}
=-\mathcal{A}\left(D^+ \rightarrow  \pi^+ \pi^0\right)_{previous}$.}
\begin{equation} \label{eq:isospin}
\frac{1}{\sqrt{2}} \mathcal{A}(D^0\rightarrow \pi^+ \pi^- ) + 
\mathcal{A}(D^0\rightarrow \pi^0 \pi^0) = \mathcal{A}(D^+\rightarrow \pi^+ \pi^0)  \, .
\end{equation}

One may ask whether there are also approximate 
sum rules which are broken by
a single matrix element only. 
By calculating the rank of the corresponding matrices of Clebsch-Gordan coefficients, we find that among the SCS decays there is exactly one such sum rule.
It is given by
the  (quasi-)triangle relation
\begin{align}
& \mathcal{A}(D^+\rightarrow \pi^+ \pi^0 ) - \frac{1}{\sqrt{2}} \mathcal{A}(D_s \rightarrow K^0 \pi^+ ) -  \nonumber\\
& \quad \mathcal{A}(D_s \rightarrow K^+ \pi^0 ) = \Sigma  \sqrt{\frac{3}{14}} B_{27}^{24_1}\,,
\label{eq:tri}
\end{align}
which generalizes Ref.~\cite{Kwong:1993ri}, where $SU(3)$ breaking by triplet matrix elements only was
considered.
Setting $B_{27}^{24_1}=0$ the sides $|A_i|$ of the triangle have to obey $|A_{\rm max}|-|A_2|-|A_3|\leq0$, {\it i.e.}, the longest of the three sides has to be smaller than the sum of the other two. This relation, if broken, would prove the necessity for  $B_{27}^{24_1}\neq0$; it is, however, fulfilled by the data, given in Table \ref{tab:data}. In the system before reparametrizations the right-hand side of Eq.~(\ref{eq:tri}) involves two matrix elements, $B_{27}^{24_1}$ and $B_{27}^{42}$,
and there is no sum rule with just one breaking matrix element.

In case the matrix element $B_{27}^{24_1}$ can be neglected and assuming MFV/SM, Eq.~(\ref{eq:tri})  allows
to extract the size of the penguin contributions by measuring  the involved branching ratios
and CP-asymmetries by the usual triangle construction. It involves a common base 
$|\mathcal{A}(D^+\rightarrow \pi^+ \pi^0|= |\mathcal{A}(D^-\rightarrow \pi^- \pi^0)|$ for the
triangle and its CP-conjugate one; the requisite weak phase $\gamma$
can  be taken from global CKM fits. A finite $B_{27}^{24_1}$ induces a correction
to this procedure of the order of the $SU(3)$-breaking, which however, is not so small.

Note that once CF and DCS modes are considered as well,  further sum rules arise, see \cite{Brod:2012ud}.

\section{SM/MFV Fits \label{sec:mfv-fit}}

We confront the $SU(3)$-analysis from the previous section to data. The relevant measurements are compiled in Table \ref{tab:data}.
The fits are carried out using the augmented lagrangian \cite{conn1991globally, birgin2008improving} and Sbplx/Subplex algorithms \cite{Rowan1990,NLopt} that are implemented in the \lq\lq{}NLopt\rq\rq{} code \cite{NLopt}. 

Our goal is to see whether $SU(3)$ gives a reasonable expansion for the full set of two-body decays of charm to pseudoscalar octet mesons,
as shown possible for $D^0\to P^+P^-$ \cite{Savage:1991wu,Pirtskhalava:2011va,Feldmann:2012js,Brod:2012ud}.
Our framework is MFV, which includes the SM. In these models CP violation is suppressed
by $\tilde \Delta$, the relative weak phase is order one, $\arg (V_{cb}^*V_{ub}/V_{cd}^*V_{ud}) =-\gamma$.

To obtain the sizable CP violation observed in SCS decays, the triplet matrix elements, $A_1^3$ or $A_8^3$, need to be sufficiently large, as they are the only ones not severely restricted by the branching ratio data. 
 In terms of the low energy effective theory, this concerns 
 penguin contractions of tree 
operators,
 the chromomagnetic dipole operator
$\bar u \sigma_{\mu \nu} G^{\mu \nu} c $, where $G^{\mu \nu}$ denotes  the gluon field strength tensor, and the QCD penguins $\bar u \gamma_\mu c \sum_q \bar q \gamma^\mu  q$,
all of which are purely $SU(3)$ triplets; we neglect the contributions from electroweak penguin operators.
On the other hand, matrix elements involving the $\rep{\bar 6}$ and $\rep{15}$ representations receive contributions from the CKM-leading tree operators only.

Given the lack of a dynamical theory for hadronic charm decays, we have to resort to order-of-magnitude arguments when judging the fit results. The penguins (triplet matrix elements) are generically expected to be suppressed by a factor of $\sim\alpha_s/\pi\sim0.1$ compared to their tree counterparts. While this estimate cannot be expected to hold literally, in the past an upper bound of one for this ratio was widely considered conservative, leading to rather strong upper limits for SM CP violation. Possible enhancements have been discussed from an early stage on \cite{Abbott:1979fw,Golden:1989qx}, and recently, {\it e.g.}, in \cite{Brod:2012ud,Franco:2012ck}. Note that the widely used analogy to the $\Delta I=1/2$ rule in kaon decays seems questionable, as the effect is expected to scale as $m_s/m_c$ \cite{Abbott:1979fw}. However, enhancements, {\it e.g.}, from rescattering cannot be excluded, but the necessary size for reaching the present central value seems a stretch, even in these analyses. Below, we introduce measures to quantify penguin enhancement for our analysis, and discuss the results obtained with present data.

\subsection{Observables vs. degrees of freedom}

The branching ratio of a decay $d$ into two pseudoscalars $P_{1,2}$ in terms of its amplitude $\mathcal{A}(d)$ is given as
\begin{align}
\mathcal{B}(D\rightarrow P_1P_2) &= \tau_D  \, \mathcal{P}(d) \,   \vert \mathcal{A}(d)\vert^2 ,
\end{align}
with the phase space  and normalization factor
\begin{align}
\mathcal{P}(d) &= \frac{ \sqrt{( m_D^2 -(m_{1} - m_{2})^2 ) ( m_D^2 - ( m_{1} + m_{2} )^2 )} }{
	 16 \, \pi \, m_D^3 } \, .
\end{align}
We take the lifetimes $\tau_{D_i}$ and masses $m_i$ of the involved mesons from  
\cite{Beringer:1900zz}.

For decays involving $D^0$ or $K^0$, the direct CP asymmetry defined in Eq.~(\ref{eq::acpdir}) is not measured directly in the experiment. The corresponding indirect  contributions are subtracted before fitting, see Appendix~\ref{app:removeMixing} for details. 

We further employ differences and sums of direct CP asymmetries of SCS decays to final states $f_{1,2}$, defined as
\begin{align}
\Delta a_{CP}^{\mathrm dir} (f_1,f_2) = a_{CP}^{\mathrm dir}(f_1)-a_{CP}^{\mathrm dir} (f_2)\, , \\
\Sigma a_{CP}^{\mathrm dir} (f_1,f_2) = a_{CP}^{\mathrm dir}(f_1)+a_{CP}^{\mathrm dir} (f_2)\,,
\end{align}
respectively. Considering $\Delta a_{CP}^{\rm dir}(f_1,f_2)$ instead of the individual asymmetries  is experimentally advantageous; furthermore, to a very good accuracy, indirect contributions cancel in the difference.

We consider now the parameter budget of the $SU(3)$ ansatz including breaking effects at leading order
versus the available experimental data. 
The 17
decay modes
correspond in principle to 26 observables, {\it i.e.}, 17 branching ratios,  
8 (direct) CP asymmetries of SCS decays, and  
the  
strong phase difference between $D^0\to \pi^+K^-$ and $D^0\to \pi^-K^+$.
Since, however, the branching ratio $\mathcal{B}(D_s \rightarrow K_L K^+)$ is not measured yet,
we are left with 25 observables.

On the parameter side within MFV/SM, there are three $SU(3)$-limit matrix elements that come with a factor of $\Sigma$ 
and two $SU(3)$-limit matrix elements that come with a factor of $\tilde\Delta$ only. Of the five resulting  
matrix elements 
one can be chosen real as we are only sensitive to differences of strong phases. In the $SU(3)$-limit there are hence nine real parameters.
Taking into account $SU(3)$-breaking, we end up with 13
independent matrix elements, see Section \ref{sec:su3break}, {\it i.e.}, 25 real parameters.

\subsection{The fate of unbroken $\mathbf{SU(3)}$}
The need for flavor-breaking in SCS decays is 
most obvious in
the large difference in the rates of the $K^+K^-$ and $\pi^+ \pi^-$ modes,  
and in the enhanced branching ratio for $D^0 \to K_S K_S$, whose contribution $\propto\Sigma$ vanishes in the $SU(3)$ limit.
Fitting in this limit CF and DCS modes only 
also returns a very large $\chi^2$, indicating sizable flavor breaking in these modes as well.
This is seen, {\it e.g.}, in
${\cal{B}}(D^0 \to K^+ \pi^-)/[\lambda^4 {\cal{B}}(D^0 \to K^- \pi^+)  ]\sim 1.5$.

Fits in unbroken $SU(3)$ using CF modes alone are not possible without additional assumptions, as the respective decay amplitudes have too much of a  linear dependence, see also the observations made in \cite{Falk:2001hx}. 
One may entertain the possibility of a 
CF-only fit by including $SU(3)$-singlets in the final states.
While this leads to additional  observables and constraints,
 it also leads to additional matrix elements. The fit presented in Ref.~\cite{Bhattacharya:2009ps} yields $\chi^2=1.79$ for 1 degree of freedom ($dof$)
at the price of an additional dynamical assumption, which effectively relates matrix elements involving singlets and octets.

\subsection{Fitting flavor-breaking \label{sec:flavorbreaking}}

Including linear $SU(3)$-breaking, we generically obtain good fits for a multitude of configurations. 
We find that a reasonable $\chi^2/dof$  requires  at least two $SU(3)$-breaking matrix elements present.
However, a fit with the assumption of triplet enhancement, as proposed in \cite{Abbott:1979fw,Kwong:1993ri}, and recently investigated in \cite{Pirtskhalava:2011va}, does not yield acceptable results ($\chi^2/dof = 8.6$), 
as no $SU(3)$ breaking enters the CF/DCS sector, see Table \ref{tab:SU3breakoctet}.
Therefore, at least one matrix element other than $B^3_{1,8}$ is needed 
to describe the data;
using, for instance, $B_1^3$ and $B_{27}^{15_2}$ we obtain $\chi^2/dof = 1.3$.
Note that a fit with $B_1^3$ to the SCS modes alone does work, $\chi^2/dof=1.0$.

To evaluate the convergence of the $SU(3)$-expansion, we  quantify the
size of the flavor-breaking with the following measures:
The size of the $SU(3)$-breaking  matrix elements, defined as
\beq \delta_X =\frac{{\rm max}_{ij}|B_i^j|}{{\rm max}(|A_{27}^{15}|, |A_{8}^{\bar 6}|, |A_8^{15}|)}  \, ,
\eeq
and the size of  the $SU(3)$-breaking amplitude, written as 
\beq
\delta_X^\prime={\rm max}_{d}\left| \frac{{\cal{A}}_{\rm X}(d)}{{\cal{A}}(d)} \right| \, . \label{eq_measure_Xprime} 
\eeq
To avoid a bias in the latter definition, 
we exclude the decay $D^0\rightarrow K_S K_S$ for which the amplitude in the $SU(3)$ limit is $\propto\tilde\Delta$.
We use both measures, as  $\delta_X$ ignores the possibility of a suppression of $SU(3)$-breaking from Clebsch-Gordan coefficients, while $\delta_X^\prime$ ignores the possibility of large cancellations.

\begin{figure}[hbt]
\begin{center}
\includegraphics[width=0.35\textwidth]{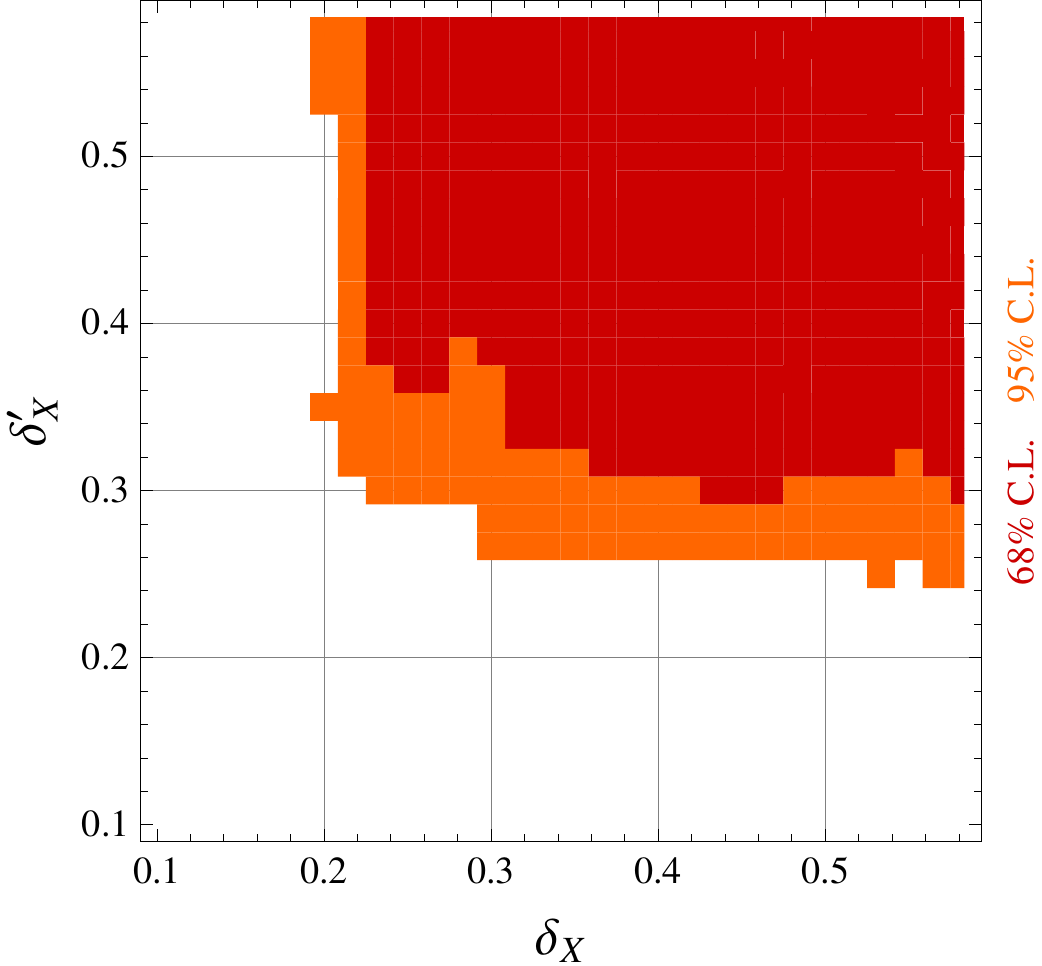}
\caption{The 68\% (dark red) and 95\% (light orange) C.L. contours in the $\delta_X$-$\delta'_X$ plane with respect to the best fit point. \label{fig_delta1delta2} }
\end{center}
\end{figure}

We find that the data can be described by an $SU(3)$-expansion with 
$\delta_{X}^{(\prime)}\lesssim 30\%$, see Fig.~\ref{fig_delta1delta2}, where we show  68\% (dark red) and 95\% (light orange) confidence level (C.L.) contours relative to the best fit point, see Section~\ref{sec:penguinenhancement}.
While solutions with larger $SU(3)$ breaking are not excluded by the data, we take this result as confirmation that the expansion works as good as could be expected. We therefore assume its validity in the following, and exclude solutions with cancellations by imposing $\delta_{X}^{(\prime)}\leq50\%$ in most fits.

Having established that with the  current data the $SU(3)$-expansion can be applied, we 
study the anatomy of the fit solutions.
With the upper bound on $\delta_X^{(\prime)}$,
we obtain a fairly good fit to the full data set with $\chi^2/dof = 1.6$ for the configuration mentioned above, which consists of only the two  matrix elements 
$B_1^3$ and $B_{27}^{15_2}$. We stress that the minimal number of $B_i^j$ is two. 
A nicer fit, however, is obtained if at least three flavor-breaking matrix elements are present. Such examples are 
the configurations with $B_1^3$, $B_8^{15_2}$, $B_{27}^{24_1}$ or
$B_1^3$, $B_{27}^{15_1}$, $B_{27}^{15_2}$, both of which give $\chi^2/dof = 1.0$. 
In the following we do not look at specific configurations with a minimal number of flavor breaking matrix elements but rather fit the full system with all $B_i^j$.

\begin{table}[hbt]
{\footnotesize
\begin{center}
\begin{tabular}{ccr}
\hline\hline
Observable            &      Measurement &  References \\\hline\hline
\multicolumn{3}{c}{SCS CP asymmetries } \\\hline
$\Delta a_{CP}^{\mathrm{dir}}(K^+K^-,\pi^+\pi^-)$         & $-0.00678 \pm 0.00147$  & \cite{Amhis:2012bh,Aubert:2007if,Aaij:2011in,Staric:2008rx, Collaboration:2012qw, ICHEP12:Ko} \\ 
$\Sigma a_{CP}^{\mathrm{dir}}(K^+K^-, \pi^+\pi^-)$        &       $+0.0014 \pm 0.0039$ &  
$^\dagger$\cite{ICHEP12:Ko, Aubert:2007if, Aaij:2011in, Aaltonen:2011se, Collaboration:2012qw}     \\
$A_{CP}(D^0 \rightarrow K_S K_S)$  & $-0.23\pm 0.19$   &   \cite{Bonvicini:2000qm}  \\
$A_{CP}(D^0 \rightarrow \pi^0 \pi^0)$   &  $+0.001\pm 0.048$ & \cite{Bonvicini:2000qm} \\
$A_{CP}(D^+ \rightarrow \pi^0 \pi^+)$ & $+0.029\pm 0.029$  & \cite{Mendez:2009aa} \\
$A_{CP}(D^+ \rightarrow K_S K^+)$    & $-0.0011\pm 0.0025$ &  \cite{CKM12:Ko, Cenci:2012uy, Mendez:2009aa, Link:2001zj} \\
$A_{CP}(D_s \rightarrow K_S \pi^+)$    & $+0.031 \pm 0.015$ & $^\dagger$\cite{Cenci:2012uy,Ko:2010ng,Mendez:2009aa} \\
$A_{CP}(D_s \rightarrow K^+ \pi^0)$ & $+0.266\pm 0.228$ &  \cite{Mendez:2009aa} \\\hline

\multicolumn{3}{c}{Indirect CP Violation} \\\hline
$a_{CP}^{\mathrm{ind}}$ & $(-0.027 \pm 0.163)\cdot 10^{-2}$ & \cite{Amhis:2012bh} \\\hline 
$\delta_L\equiv 2 \mathrm{Re}(\varepsilon)/(1+\vert\varepsilon\vert^2)$   & $(3.32 \pm 0.06) \cdot 10^{-3}$ & \cite{Beringer:1900zz}   \\\hline
\multicolumn{3}{c}{$K^+ \pi^-$ strong phase difference} \\\hline
$\delta_{K \pi}$    &  $21.4^{\circ}  \pm 10.4^{\circ}$ &  $^\ddag$\cite{Amhis:2012bh}  \\\hline
\multicolumn{3}{c}{SCS branching ratios} \\\hline
$\mathcal{B}(D^0\rightarrow K^+ K^-  )   $   & $   \left(3.96\pm  0.08\right)\cdot10^{-3} $ &  \cite{Beringer:1900zz}   \\
$\mathcal{B}(D^0\rightarrow \pi^+ \pi^-) $   & $   \left( 1.401\pm  0.027\right)\cdot10^{-3} $ & \cite{Beringer:1900zz}  \\
$\mathcal{B}(D^0\rightarrow K_S K_S  )   $   & $   \left(0.17\pm  0.04\right)\cdot10^{-3} $ & \cite{Beringer:1900zz}  \\
$\mathcal{B}(D^0\rightarrow \pi^0 \pi^0) $   & $   \left(0.80\pm  0.05\right)\cdot10^{-3} $ &  \cite{Beringer:1900zz} \\
$\mathcal{B}(D^+\rightarrow \pi^0 \pi^+) $   & $   \left(1.19\pm  0.06\right)\cdot10^{-3} $ &  \cite{Beringer:1900zz} \\
$\mathcal{B}(D^+\rightarrow K_S K^+  )   $   & $   \left(2.83\pm  0.16\right)\cdot10^{-3} $ &  \cite{Beringer:1900zz} \\
$\mathcal{B}(D_s\rightarrow K_S \pi^+ )  $   & $   \left(1.21\pm  0.08\right)\cdot10^{-3} $ &  \cite{Beringer:1900zz} \\
$\mathcal{B}(D_s\rightarrow K^+ \pi^0 )  $   & $   \left(0.62\pm  0.21\right) \cdot10^{-3} $ &  \cite{Beringer:1900zz} \\\hline
\multicolumn{3}{c}{ CF$^*$ branching ratios} \\\hline
$\mathcal{B}(D^0\rightarrow K^- \pi^+ )  $   & $   \left(3.88\pm  0.05\right)\cdot10^{-2} $  &  \cite{Beringer:1900zz}\\
$\mathcal{B}(D^0\rightarrow K_S \pi^0 )  $   & $   \left(1.19\pm  0.04\right)\cdot10^{-2}  $ &  \cite{Beringer:1900zz} \\
$\mathcal{B}(D^0\rightarrow K_L \pi^0 )  $   & $   \left(1.00\pm  0.07\right)\cdot10^{-2}  $ &   \cite{Beringer:1900zz}\\
$\mathcal{B}(D^+\rightarrow K_S \pi^+ )  $   & $   \left(1.47\pm  0.07\right)\cdot10^{-2}  $ &  \cite{Beringer:1900zz} \\
$\mathcal{B}(D^+\rightarrow K_L \pi^+ )  $   & $   \left(1.46\pm  0.05\right)\cdot10^{-2}  $ &  \cite{Beringer:1900zz} \\
$\mathcal{B}(D_s\rightarrow K_S K^+  )   $   & $   \left(1.45\pm  0.05\right)\cdot10^{-2}  $ & $^\dagger$\cite{Beringer:1900zz,ICHEP12:Wang} \\\hline
\multicolumn{3}{c}{DCS branching ratios} \\\hline
$\mathcal{B}(D^0\rightarrow K^+ \pi^- )  $   & $   \left(1.47\pm  0.07\right)\cdot10^{-4}  $ &  \cite{Beringer:1900zz} \\
$\mathcal{B}(D^+\rightarrow K^+ \pi^0 )  $   & $   \left(1.83\pm  0.26\right)\cdot10^{-4}  $ &  \cite{Beringer:1900zz} \\\hline\hline
\end{tabular}
\end{center}
\caption{The observables and the data for indirect CP violation used in this work, see
 Appendix  \ref{app:removeMixing} for removal of effects from charm and kaon mixing.
$^\dagger$The measurement quoted corresponds to our average. Systematic and statistical uncertainties are added in quadrature.
$^\ddag$Our symmetrization of uncertainties. $^*$Modes into $K_{S,L}$ assigned to CF decays.
\label{tab:data}}
}
\end{table}

\subsection{Penguin enhancement \label{sec:penguinenhancement}}
Next we turn to analyze the penguin enhancement. One possible definition is the following ratio:
\beq
\delta_3 =\frac{{\rm max}(|A_1^3|, |A_8^3|)}{{\rm max}(|A_{27}^{15}|, |A_{8}^{\bar 6}|, |A_8^{15}|)} \, .
\eeq
In order to inspect also here the possibility of cancellations, we 
further define the penguin-to-tree fraction of a specific decay $d$ as
\beq
\delta_3^\prime(d) = 
\left|\frac{c_{d;1\,3}A_1^3+c_{d;8\,3}A_8^3}{c_{d;27\,15}A_{27}^{15}+c_{d;8\,\bar6}A_8^{\bar 6}+c_{d;8\,15}A_8^{15}}\right|\,,
\eeq
and its maximum
\beq
\delta_3^\prime =  {\rm max}_d \,  \delta_3^\prime(d)
\,  . \label{eq_delta3prime}
\ee
Here, for the same reason as in the case of $\delta'_X$, the decay $D^0\rightarrow K_S K_S$ is not taken into account.

\begin{figure}[hbt]
\begin{center}
\includegraphics[width=0.35\textwidth]{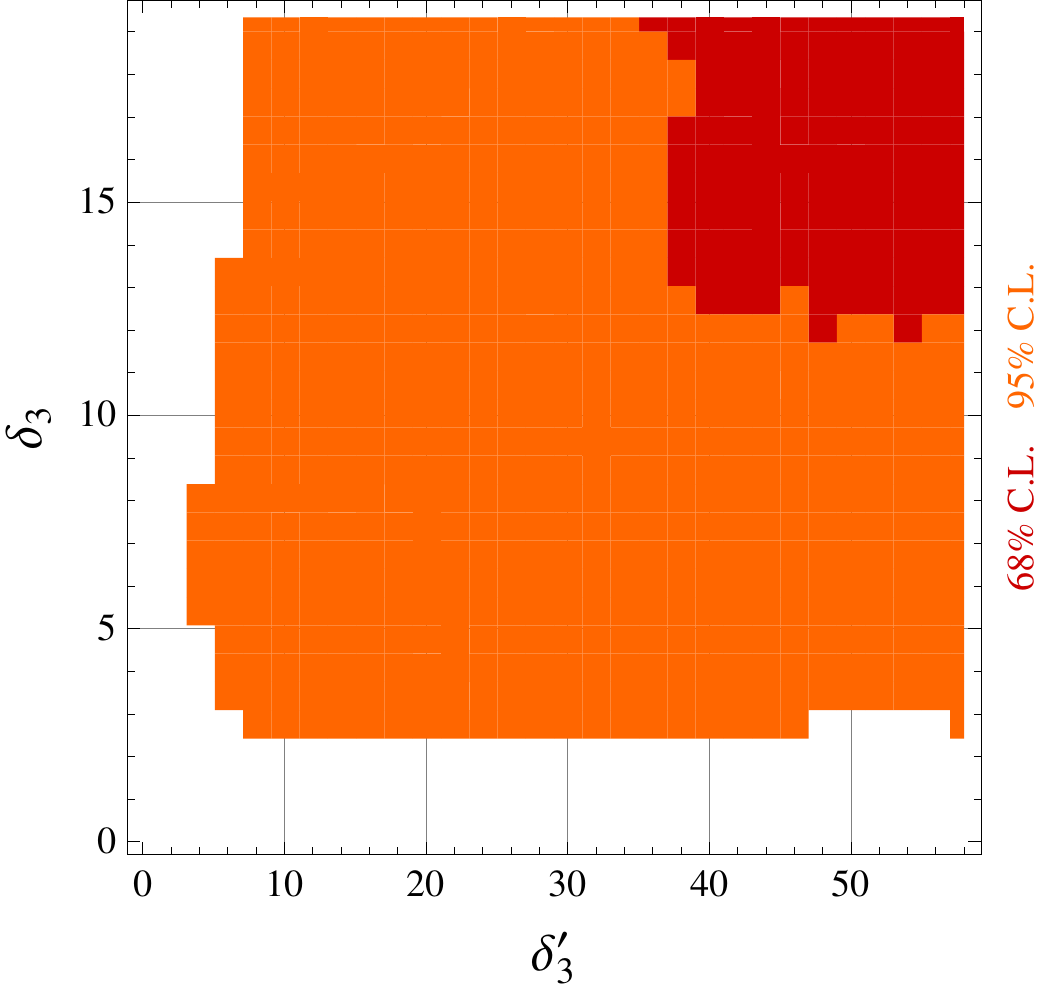}
\end{center}
\caption{The 68\% (dark red) and 95\% (light orange) C.L. contours in the $\delta'_3$-$\delta_3$ plane with respect to the best fit point.
\label{fig_delta3primedelta3}  }
\end{figure}

\begin{figure}[hbt]
\begin{center}
\includegraphics[width=0.35\textwidth]{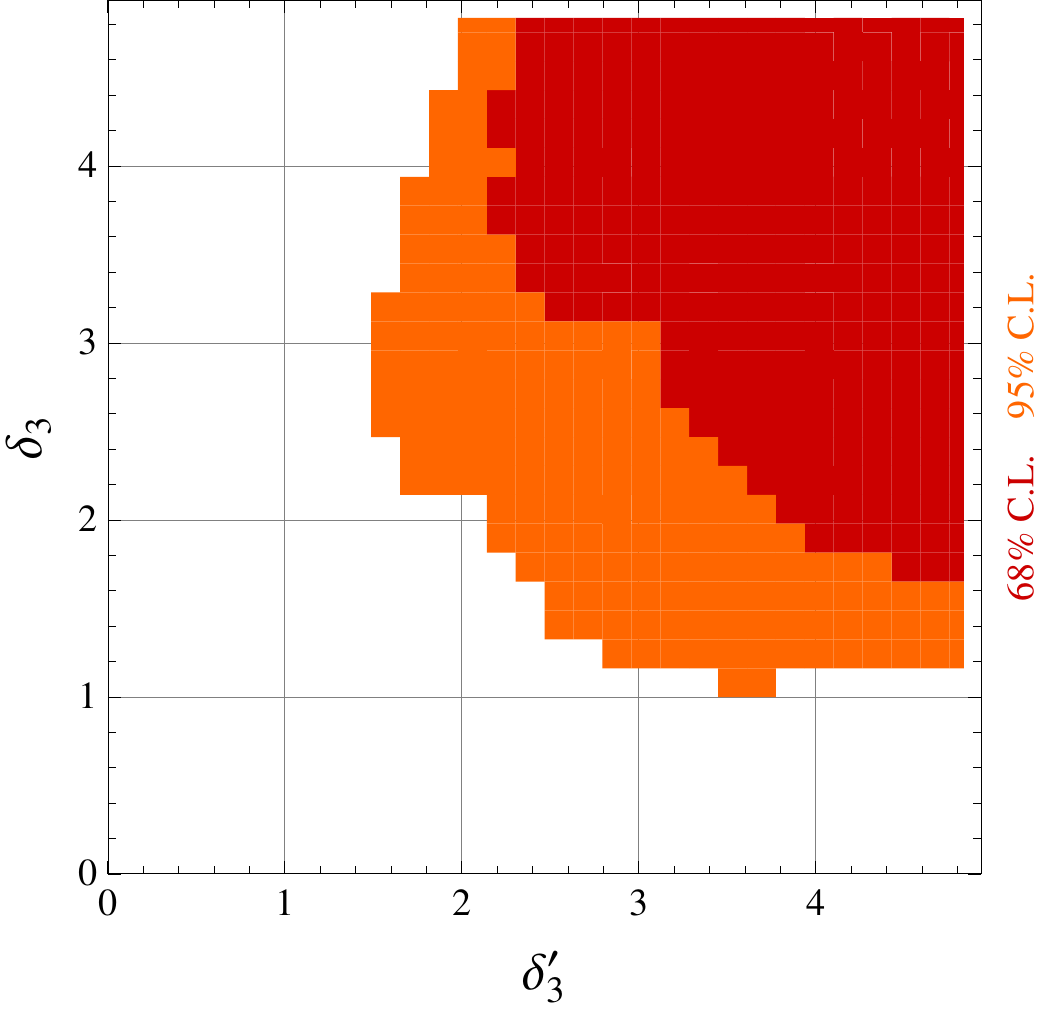}
\end{center}
\caption{The same as in Fig.~\ref{fig_delta3primedelta3} 
without the data on
$A_{CP}(D^0\rightarrow K_S K_S)$, $A_{CP}(D_s\rightarrow K_S \pi^+)$ and
$A_{CP}(D_s\rightarrow K^+ \pi^0)$.
\label{fig_delta3primedelta3_wo_obs}  }
\end{figure}

At the best fit point we obtain $\chi^2 =1.0$ for 25 real fit parameters and an equal number of observables, obtained without a constraint on $\delta_X^{(\prime)}$. The residual $\chi^2$ is caused  
by $A_{CP}(D^+\rightarrow \pi^+\pi^0)$, which is measured nonzero at $1\sigma$, but impossible to accommodate in the SM
as it vanishes therein. 
We observe that $\delta_3^{(\prime)}$ is driven to huge values of $\mathcal{O}(100)$ in the fit.  The reason for this strong enhancement lies not so much in the measurement for $\Delta a_{CP}^{\rm dir}$, but is due to the CP asymmetries $A_{CP}(D^0\rightarrow K_S K_S)$, $A_{CP}(D_s\rightarrow K_S \pi^+)$, and $A_{CP}(D_s\rightarrow K^+ \pi^0)$, which presently have very large central values, 
see Table \ref{tab:data}. Their uncertainties are very large as well, rendering the effect insignificant for each single measurement. Together however, while still allowing for solutions with $\delta_3^{(\prime)}\sim 5$ at $95\%$~C.L., they shift the $68\%$~C.L. region to very large values. Numerically, we obtain $\chi^2=1.9$ for $\delta_3\leq30$, and $\chi^2=3.6$ for $\delta_3\leq10$, $\delta_3$ in both cases saturating the bound. The latter value is slightly increased to $\chi^2=4.5$ when additionally imposing $\delta_X^{(\prime)}\leq50\%$, indicating a very small correlation between the two measures. These observations are illustrated in 
Figs.~\ref{fig_delta3primedelta3} and~\ref{fig_delta3primedelta3_wo_obs}. In the latter we show the influence of the largish measured CP asymmetries explicitly by excluding them from the fit. As a result, values of $\delta_3^{(\prime)}\sim 3$ become allowed at $68\%$~C.L., consistent with Refs.~\cite{Isidori:2011qw,Feldmann:2012js,Brod:2012ud}, where 
these asymmetries 
have not been taken into account either.

Closer inspection exhibits common features among the presented fits: 
The penguin matrix elements 
tend to be largely enhanced; there are
large hierarchies between $\delta^\prime_3(d)$, depending on whether 
both triplet matrix elements $A_1^3$ and $A_8^3$ are present, 
as is the case in the decays $D^0\rightarrow K^+ K^-$, $D^0\rightarrow \pi^+ \pi^- $ and 
$D^0\rightarrow \pi^0 \pi^0$, or not.  In the latter modes  indeed cancellations take place, 
yielding a smaller penguin amplitude with $\delta_3^\prime(d) \ll \delta_3$, and typically
$\delta_3^\prime(d)\sim {\cal{O}}(5)$.

The maximum value of $\delta^\prime_3(d)$, on the other hand,  usually comes from the decays where only  $A_8^3$ is present, {\it i.e.}, $D^+\rightarrow K_S K^+$, $D_s\rightarrow K_S \pi^+$ or $D_s\rightarrow K^+ \pi^0$. 
Since furthermore typically $ {\rm max}(|A_{27}^{15}|, |A_{8}^{\bar 6}|, |A_8^{15}|) = |A_8^{15}|$,
and the ratio of the 
Clebsch-Gordan coefficients  of $A_8^3$ and  $A_8^{15}$ in the latter modes equals $\frac{3\sqrt{5}}{2} \sim 3$, we understand why 
generically $\delta'_3 \sim 3 \cdot \delta_3$. 

Taking into account the full dataset, we therefore find indications of even stronger enhanced penguin amplitudes than previous analyses relying on $\Delta a_{CP}^{\rm dir}$ only. 
On the other hand, given the present uncertainties, the large central values responsible for this additional enhancement might be assigned to experimental fluctuations. As the enhancement indicated by $\Delta a_{CP}^{\rm dir}$ is already very difficult to explain within the SM, any additional enhancement challenges this option further, making more precise measurements of the corresponding CP asymmetries extremely important.

\section{Patterns of New Physics \label{sec:pattern}}

In the absence of a dynamical theory or further input we still cannot make a definite statement on 
whether the enhanced penguins observed in Section~\ref{sec:mfv-fit} stem from physics within or beyond the SM.
Note that knowing the 
weak phase $\gamma$ precisely  in MFV/SM is of no help as
there is an approximate parametrization invariance: because of the smallness of $|\tilde \Delta|$, the rate measurements are not sensitive to the  terms $\mathcal{O}({\rm Re}\tilde\Delta,|\tilde\Delta|^2)$. The CP asymmetries are proportional to ${\rm Im}\tilde\Delta$, however, they always involve unknown matrix elements not constrained by the rates. 
As a result, a shift in ${\rm Im}\tilde\Delta$ ({\it e.g.}, a different weak phase) can in the fit always be absorbed by a shift in the magnitude of the respective matrix elements, as long as extreme values are avoided. Therefore, we set in the following $\tilde\Delta\equiv\tilde\Delta_{SM}$, even when the NP model does not require a specific value. The potential enhancement factors from the Wilson coefficients are then identified by enhanced NP matrix elements, where we assume the absence of very large enhancement factors from QCD. Furthermore, we assume the validity of the $SU(3)$ expansion, using again the upper limit $\delta_X^{(\prime)}\leq 50\%$.

To investigate the interplay of NP and $SU(3)$ breaking, 
we study the following $SU(3)$-limit relations involving CP asymmetries:
\begin{align} 
\label{eq:kkpipi0} \frac{\Gamma(D^0\to K^+K^-)}{\Gamma(D^0\to \pi^+\pi^-)} &= -\frac{a_{CP}^{\rm dir}(D^0\to \pi^+\pi^-)}{a_{CP}^{\rm dir}(D^0\to K^+K^-)}\,,\\
\label{eq:k0k+0} \frac{\Gamma(D^+\to \bar{K}^0K^+)}{\Gamma(D^+_s\to K^0\pi^+ )} &= -\frac{a_{CP}^{\rm dir}(D^+_s\to K^0\pi^+ )}{a_{CP}^{\rm dir}(D^+\to \bar{K}^0K^+)}\,,\\
a_{CP}^{\mathrm dir} (D^0 \to K^0 \bar K^0)&=0 \, , \label{eq:k0k0b}\\
a_{CP}^{\mathrm dir} (D^+ \to \pi^+ \pi^0)&=0\, .\label{eq:piplpi0}
\end{align}
The first three relations follow in the $U$-spin limit, while the last one is an isospin relation. All relations
can be obtained by inspecting, for instance, Table~\ref{tab::SU3limitoctet}.
Eqs.~(\ref{eq:kkpipi0}) and~(\ref{eq:k0k+0}) can be rewritten as 
\begin{align}
\label{eq:kkpipi}\Sigma a_{CP}^{\mathrm dir} (K^+ K^-,\pi^+ \pi^-) &= 0\,,\\
\label{eq:dpds}\Sigma a_{CP}^{\mathrm dir} (\bar{K}^0K^+,K^0\pi^+) &= 0\,,
\end{align} 
with corrections of the order $\mathcal{O}({\rm Re}\tilde\Delta {\rm Im}\tilde\Delta)$, which are completely negligible compared to $SU(3)$-breaking corrections. In addition, the $SU(3)$-breaking representations $\rep{\bar 6}$ and $\rep{24}$ do not yield corrections to these relations, either.

The $SU(3)$-expansion allows to quantify corrections to the above relations.
Further deviations 
then indicate the presence of  $U$- and isospin changing interactions
beyond the SM, models of which are discussed in the next section.

\subsection{New physics scenarios}

We recall the $SU(3)$-limit Hamiltonian $\mathcal{H}^{\rm SCS}_{\rm SM}$ 
for SCS decays within MFV/SM:
\begin{align}
\mathcal{H}^{\rm SCS}_{\rm SM}&= \mathcal{H}^{\rm SCS}_{\Delta,SM}
+ \mathcal{H}^{\rm SCS}_{\Sigma}\,, \nonumber \\
\mathcal{H}^{\rm SCS}_{\Sigma}&= \Sigma\left(-\frac{1}{\sqrt{3}}\rep{15}_{3/2}+\sqrt{\frac{2}{3}}\rep{15}_{1/2}-\rep{\bar{6}}_{1/2}\right) \,,
\end{align}
and $\mathcal{H}^{\rm SCS}_{\Delta,SM}$ involving the representations~$\rep{3}$ and $\rep{15}$, which, however, do not contribute significantly
due to our assumption of ``well-behaved'' hadronic matrix elements, and are therefore neglected in the following.
Here and in the following the subscript to the representation labels the  shift in total isospin $\Delta I$.

We discuss scenarios $\mathcal{H}^{\rm SCS}_{NP}\equiv\mathcal{H}^{\rm SCS}_{reps}+\mathcal{H}^{\rm SCS}_\Sigma$, which can be classified according to the contributing $SU(3)$-representations. We use $\Delta \equiv \tilde \Delta \Sigma=(V_{cs}^*V_{us}+V_{cd}^*V_{ud})/2$.

Very similar to the SM/MFV are models based on new physics in the $\rep{3}$ alone (``triplet model''), 
\begin{align} \label{eq:H3}
\mathcal{H}^{\rm SCS}_3&=\Delta \sqrt{\frac{3}{2}}\rep{3}_{1/2}^{\mathrm{NP}} \, .
\end{align}
At the quark level, the corresponding operators are the chromomagnetic dipole operator
and the QCD penguin operators. They have been discussed in supersymmetric  \cite{Grossman:2006jg,Giudice:2012qq,Hiller:2012wf} and extradimensional models  \cite{DaRold:2012sz} in the context of CP violation in charm.

We further consider a $\rep{3}+ \rep{15}$ interaction
\begin{align}
\mathcal{H}^{\rm SCS}_{3+15}&=  \Delta\left( \rep{15}^{\mathrm{NP}}_{3/2}+\frac{1}{\sqrt{2}}\rep{15}^{\mathrm{NP}}_{1/2}+\sqrt{\frac{3}{2}}\rep{3}^{\mathrm{NP}}_{1/2}\right)\,,
\end{align}
which arises from an operator with flavor structure $\bar u c \bar u u$ (``HN model''). Those have been
investigated recently in a peculiar variant of a 2-Higgs doublet model (2HDM)  that links the top sector to charm  \cite{Hochberg:2011ru}.
While the flavor structure is identical to the corresponding SM tree operator, it can have a different Dirac structure. As a result, the matrix elements of the $\rep{15}$-representation are independent of those appearing with a coefficient $\Sigma$, and therefore not constrained by the data on the rates.

Finally, we allow for $\rep{3}+ \rep{15}+\rep{\bar 6}$ terms (``$\Delta U=1$ model''),
\begin{align} \label{eq:H3156}
\mathcal{H}^{\mathrm{SCS}}_{3+\bar6 +15} &= \Delta \left(
                 \sqrt{\frac{3}{2}} \mathbf{15}^{\mathrm{NP}}_{1/2}
                - \mathbf{\bar{6}}^{\mathrm{NP}}_{1/2}
		-\sqrt{\frac{3}{2}} \mathbf{3}^{\mathrm{NP}}_{1/2} 
\right) \,  .
\end{align}
The corresponding operators have flavor content $\bar s c \bar u s$. This structure may  arise from tree level scalar exchanges as in 2HDMs or a color octet, see \cite{Altmannshofer:2012ur} for a list. The appearance of a third representation carrying a weak phase different from the leading contributions implies significantly less correlation between different CP violating observables. This makes this scenario especially difficult to identify in patterns.

All models contribute to SCS decays only.
We do not consider potential constraints on the scenarios with NP from 4-Fermi operators by $D^0 -\bar D^0$ mixing and $\epsilon^\prime/\epsilon$ \cite{Isidori:2011qw}.

As QCD preserves $SU(3)$-flavor, the irreducible representations form subsets which renormalize only among themselves \cite{Golden:1989qx}. 
While such effects often have numerical relevance, they do not for our analysis,
as we fit matrix elements rather then calculating them.
We still ask whether the $SU(3)$-anatomy of the NP models considered is radiatively stable.

A NP contribution in the $\rep{3}$ does not ``spread out'' by mixing onto other representations.
The situation for the scalar operators $\bar s_R c_L \bar u_R s_L$
is likewise simple: They mix, together with their scalar and tensor color-flipped partners with the same flavor content  at leading order  only among themselves. The mixing onto the dipole operators vanishes for $m_s=m_d=0$ \cite{Buras:2000if,Altmannshofer:2012ur}. 
The total contribution to the amplitude therefore remains $\rep{3}+ \rep{15}+\rep{6}$.

The scalar  $\bar u_R c_L \bar u_L u_R$  mixes among itself, color-flipped partners and onto chirality-flipped QCD-penguins; hence,
renormalization group (RG) running modifies the weight between the $\rep{3}$ and the $\rep{15}$ set at the NP
scale by Clebsch-Gordan coefficients. However, an explicit calculation of the leading order
running between the weak and the charm scale shows that this effect is $\lesssim 7 \%$, and can be safely neglected for the purpose of our analysis. Anomalous dimensions can be taken from, {\it  e.g.}, \cite{Hiller:2003js}.

The RG stability depends in general on the Dirac structure. We use in the following the corresponding $SU(3)$-classifications,
but have in mind the model examples where RG effects can be neglected.

\subsection{Patterns}
In this section we investigate if and how the different NP scenarios can be distinguished from each other and MFV/SM. To that end, we perform fits to the full data set in each scenario and look for specific correlations.

Given the fact that only one combination of CP asymmetries is measured significantly non-zero so far, the differentiation of models with present data is extremely difficult. We therefore consider in our analysis in addition a future data set, see Table~\ref{tab:future} and Appendix~\ref{app:futdata} for details.

Note that within our framework we are not able to distinguish a NP $\rep{3}$ from the SM. 
A recent idea to
resolve this is to measure radiative charm decays, where the sensitivity to enhanced dipole operators is enhanced w.r.t hadronic decays \cite{Isidori:2012yx}.

We start by analyzing some generic features of  Eqs.~(\ref{eq:kkpipi0})-(\ref{eq:piplpi0}). The most clearcut relation is the isospin one, Eq.~(\ref{eq:piplpi0}); it holds even in the presence of $SU(3)$ breaking, as long as no  operator with a $\rep{15}_{3/2}$ representation different from the SM one is present, see also \cite{Grossman:2012eb}. Therefore, it serves as an extremely clean ``smoking gun'' signal for the HN model.

The unique feature of Eq.~(\ref{eq:k0k0b}) is the vanishing of the CKM-leading part of the $D^0\to K_SK_S$ amplitude in the $SU(3)$ limit. The size of the rate is determined by the $SU(3)$-breaking contribution; the corresponding CP asymmetry is therefore enhanced at  $\mathcal{O}(\tilde\Delta/\delta_X^{(\prime)})$. This enhancement might roughly be estimated as 
\begin{align}\label{eq:KsKs}
\frac{a_{CP}^{\rm dir}(D^0\to K^0\bar K^0)}{a_{CP}^{\rm dir}(D^0\to K^+K^-)}&\sim\sqrt{\frac{BR(D^0\to K^+K^-)}{BR(D^0\to K^0\bar K^0)}}\sim3\,,
\end{align}
implying $a_{CP}^{\rm dir}(D^0\to K^0\bar K^0)\sim1\%$ for $a_{CP}^{\rm dir}(D^0\to K^+K^-)\sim \Delta a_{CP}^{\rm dir}/2$.
While this might be considered an upper limit for the SM, as the asymmetry in the charged final state is already larger than expected therein, further enhancements are possible within the NP scenarios. The sole difference between the NP models is how strongly $a_{CP}^{\rm dir}(D^0\to K_SK_S)$ is correlated to other CP asymmetries. In the MFV/SM and the triplet scenario it is determined by the triplet matrix elements alone, while in the other two NP scenarios the $\rep{15}$ can give a significant contribution as well.

\begin{table}[hbt]
\begin{center}
\begin{tabular}{cl}
\hline\hline
Observable   & Future data   \\\hline\hline
\multicolumn{2}{c}{SCS CP asymmetries } \\\hline
$\Delta a_{CP}^{\mathrm{dir}}(K^+K^-,\pi^+\pi^-)$         & $-0.007\pm 0.0005$   \\\hline
$\Sigma a_{CP}^{\mathrm{dir}}(K^+K^-, \pi^+\pi^-)$        & $-0.006\pm 0.0007\phantom{0}$   \\\hline
$a_{CP}^{\mathrm{dir}}(D^+ \rightarrow K_S K^+)$     & $-0.003\pm 0.0005\phantom{0}$   \\\hline

$a_{CP}^{\mathrm{dir}}(D_s \rightarrow K_S \pi^+)$   & $\phantom{-}0.0 \pm 0.0005$    \\\hline
$a_{CP}^{\mathrm{dir}}(D_s \rightarrow K^+ \pi^0)$ & $\phantom{-}0.05\pm 0.0005$    \\\hline
\multicolumn{2}{c}{$K^+ \pi^-$ strong phase difference} \\\hline
$\delta_{K \pi}$    &  $21.4^{\circ}  \pm 3.8^{\circ}$ \\\hline\hline 
\end{tabular}
\end{center}
\caption{Future data, all other values as in Table~\ref{tab:data}. 
The central values of the single CP asymmetries that correspond to $\Delta a_{CP}^{\mathrm{dir}}$ and 
$\Sigma a_{CP}^{\mathrm{dir}}$ are $a_{CP}^{\mathrm{dir}}(D^0\rightarrow K^+K^-) = -0.0065$ and 
$a_{CP}^{\mathrm{dir}}(D^0\rightarrow \pi^+ \pi^-) = 0.0005$.
 \label{tab:future}}
\end{table}

The remaining two relations Eqs.~(\ref{eq:kkpipi})-(\ref{eq:dpds}) are broken differently in different NP models.  In MFV/SM, as well as in the triplet and HN scenarios, the sum of the CP asymmetries receives two contributions at $\mathcal{O}(\delta_X)$: one is proportional to $\Delta a_{CP}^{\rm dir}$, driven by the relative rate difference of the two modes, 
{\it i.e.}, the contribution simply  stems from the different normalization of the two CP asymmetries. The other contribution stems from the interference of the $SU(3)$ breaking part of the amplitude with the part $\propto \tilde\Delta$.  The sign of the total contribution can not trivially be extracted from the ratio of the rates. 

In addition to these $SU(3)$-breaking contributions, in the $\Delta U=1$ model the relations Eqs.~(\ref{eq:kkpipi})-(\ref{eq:dpds}) are broken at $\mathcal{O}(1)$ by NP. The reason for that is that in this model there is no discrete $U$-spin symmetry of the Hamiltonian under exchanging all down and strange quarks anymore. Therefore, generically the contributions of the $\Delta U=1$ model are expected to be larger than in the other NP scenarios. 

Not unexpectedly, all scenarios fit the current data well, with a minimal $\chi^2\sim1$. The main difference lies in the interpretation of the enhancements of the various matrix elements. In addition, as mentioned above, the HN model has the advantage of being able to explain the CP asymmetry in $D^0\to \pi^0\pi^+$ as well, leading to $\chi^2\sim0$. Excluding this measurement, all scenarios have a vanishing minimal $\chi^2$. 

In the HN model we get good fits with 
$A_{27}^{15,\mathrm{NP}} / A_8^{15} \sim 10$ which is essentially determined by the 1$\sigma$ measurement of $a_{CP}^{\mathrm{dir}}(D^+ \rightarrow \pi^0 \pi^+)$, because the $A_{27}^{15,\mathrm{NP}}$ is the only contributing NP matrix element to this mode. As generic size of the $\bar u_R c_L \bar u_L u_R$ to  tree enhancement we obtain
\begin{align}
{G}^{SU(3)} &= \frac{ 
\frac{3 \sqrt{3/2}}{10} A_{27}^{15, NP} 
}{
\frac{2}{5 \sqrt{2}} A_8^{15}
 } \sim 13\,, 
\end{align}
where we accounted for the Clebsch-Gordan coefficients. This value is in agreement with \cite{Hochberg:2011ru}.
It also fits the recent results on the forward-backward $t\bar{t}$ production asymmetry from the full data set from the CDF experiment \cite{Aaltonen:2012it}. The related bound from the LHC on the charge asymmetry 
$A_C$ \cite{ATLAS:2012an,CMS-PAS-TOP-11-014} can ``presumably'' be evaded by the HN model \cite{Drobnak:2012rb}.

We learn that with present data a clear separation between different NP models is not possible. There are two paths to obtain a clearer picture. Either we gain insights in the dynamics of $SU(3)$ breaking, which would, {\it e.g.}, allow us to identify certain matrix elements as leading in the breaking, or we wait for more precise data to see if the patterns of NP discussed above become significant.

To demonstrate how an improved understanding of $SU(3)$ breaking would improve our fits, we consider one of the scenarios mentioned above with only three additional matrix elements, $B_1^3,B_8^{15_2}$, and $B_{27}^{24_1}$. As for this case MFV/SM already fits the data well, so do the NP scenarios. We obtain $\chi^2/dof=10/10$ in MFV/SM and the triplet model, $\chi^2/dof=3.2/6$ for the HN model, and $\chi^2/dof=5.4/4$ for the $\Delta U=1$ model. The slightly worse result for the $\Delta U=1$ model is due to the fact that the additional degrees of freedom introduced are not necessary to fit the CP asymmetries, given that only $\Delta a_{CP}^{\rm dir}$ is measured significantly different from zero. 

\begin{figure}[hbt]
\begin{center}
\includegraphics[width=0.35\textwidth]{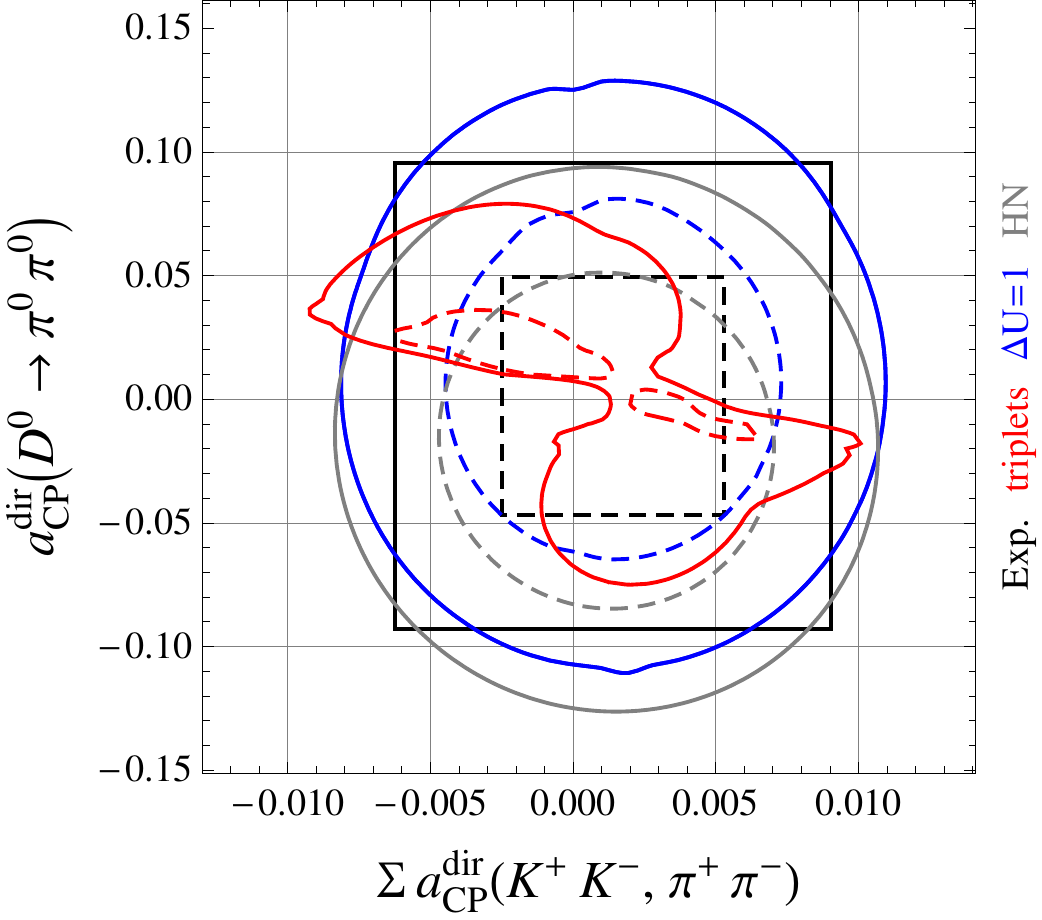}
\caption{95\% (solid) and 68\% (dashed) C.L. contour lines for the current data with only three breaking matrix elements ($B_1^3$, $B_8^{15_2}$, $B_{27}^{24_1}$) with $\delta_X^{(\prime)} \leq 50\%$. \label{fig:sum-pi0pi0-a-minimal}} 
\end{center}
\end{figure}

We find that for such an $SU(3)$-breaking scenario, the different NP models start to imply different patterns. This is illustrated in Fig.~\ref{fig:sum-pi0pi0-a-minimal}, where we observe a correlation between the signs of $a_{CP}^{\rm dir}(D^0\to\pi^0\pi^0)$ and $\sum a_{CP}^{\rm dir}(K^+K^-,\pi^+\pi^-)$ in the triplet model. 
The capability to differentiate between the different models will improve significantly with future data. With present data, however, it is already possible to exclude many scenarios for $SU(3)$ breaking, as the exclusion of the pure triplet enhancement  demonstrated in Section~\ref{sec:flavorbreaking}.

\begin{figure}[hbt]
\begin{center}
\includegraphics[width=0.35\textwidth]{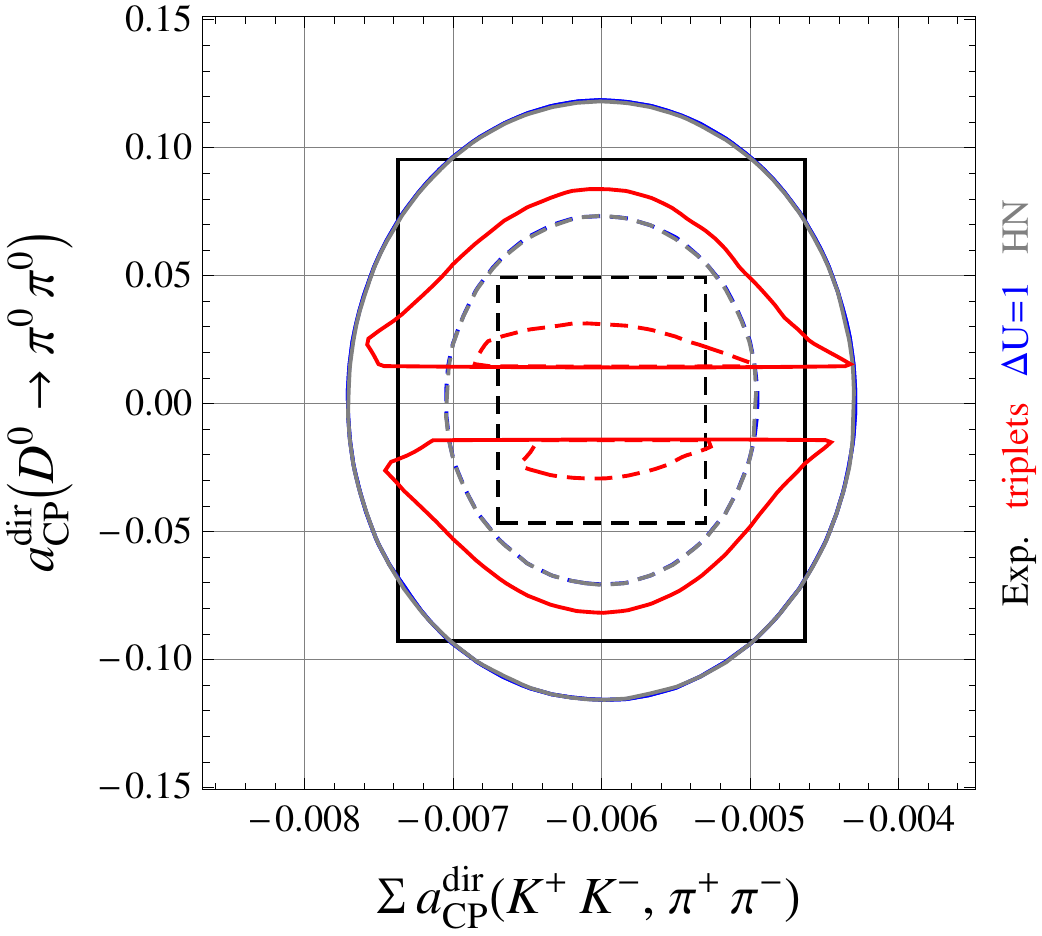}
\caption{ 95\% (solid) and 68\% (dashed) C.L. contour lines for the future data set with $\delta_X^{(\prime)} \leq 50\%$. Note that the contour for the $\Delta U=1$ model lies underneath the one of the HN model.
\label{fig:sum-pi0pi0-future-deltaXprime-bound} } 
\end{center}
\end{figure}

The future data scenario is designed having in mind the $\Delta U=1$ model being realized. The goal is to determine whether this model can be distinguished from the others by $D\to PP$ data, despite its many CP violating contributions. We find that all NP models remain capable of fitting the future data set well, despite its somewhat specific construction, with  $\chi^2_{\rm min}\sim0$ for the HN model and $\chi^2_{\rm min}\sim1$ for the others.  In Fig.~\ref{fig:sum-pi0pi0-future-deltaXprime-bound}, we show exemplarily correlations for the various models as in Fig.~\ref{fig:sum-pi0pi0-a-minimal}. 
While for the shown observables the $\Delta U=1$ model cannot be distinguished from the HN model, in MFV/SM and the triplet model a clear prediction of a sizable CP asymmetry in $D^0\to\pi^0\pi^0$ emerges. This serves as a confirmation that in principle the various scenarios are distinguishable within our framework. However, as is obvious comparing Figs.~\ref{fig:sum-pi0pi0-a-minimal} and \ref{fig:sum-pi0pi0-future-deltaXprime-bound}, further dynamical input on $SU(3)$ breaking would facilitate this task enormously. 

\section{Conclusions \label{sec:end}}

We performed a comprehensive $SU(3)$-flavor analysis using the complete set of data
on two-body decays of $D$-mesons to pseudoscalar octet mesons.
The results are based on plain $SU(3)$; in particular, we did not make any assumptions about
 decay topologies nor assumed factorizable $SU(3)$-breaking.
We find
\begin{itemize}
\item[\it i] The $SU(3)$-expansion can describe the CF, SCS and DCS data set with 
breaking of ${\cal{O}}(30) \%$.
\item[\it ii] $SU(3)$-breaking matrix elements involving  higher representations  cannot be neglected.
\item[\it iii] Current data imply significantly enhanced penguin  
matrix elements with respect to the
non-triplet ones, see Figure \ref{fig_delta3primedelta3}.
If the measured largish CP asymmetries in $D^0\rightarrow K_S K_S$,  
$D_s\rightarrow K_S \pi^+$ and $D_s\rightarrow K^+ \pi^0$ decays are  
excluded from the fit the
penguin enhancement is shifted to smaller but  still significantly  
enhanced values at ${\cal{O}}(2-5)$, see
Figure \ref{fig_delta3primedelta3_wo_obs}, favoring interpretations   
within NP models.
Improved data could clarify how large the penguin enhancement actually  
needs to be.
\end{itemize}

The model-independence of the $SU(3)$-expansion leads to a large number of $SU(3)$-breaking matrix elements,
and the outcome of the fits currently does not
 allow an unambiguous interpretation regarding the underlying electroweak physics. 
In more minimal scenarios, where $SU(3)$-breaking is limited to
 a smaller number of matrix elements, clearer patterns exist, see Figure~\ref{fig:sum-pi0pi0-a-minimal}. In addition, it is possible already with present data to differentiate between various $SU(3)$-breaking structures.

Keeping all matrix elements, the following measurements of direct CP violation are found to be informative for discriminating scenarios:
\begin{itemize}
\item[{\it iv}] A breakdown at $\mathcal{O}(1)$ of the relations Eqs.~(\ref{eq:kkpipi}) and (\ref{eq:dpds}) between
$D^0 \to K^+ K^-$ versus $D^0 \to \pi^+ \pi^-$ and
$D_s \to K_S \pi^+$ versus $D^+ \to K_S K^+$ 
would indicate $\Delta U \neq 0$ new physics. An observation would
support the $\rep{3}+\rep{\bar 6}+\rep{15}$ Hamiltonian.
 \item[{\it v}] An observation of a finite $a_{CP}^{\mathrm dir} (D^+ \to \pi^+ \pi^0)$ would signal $\Delta I =3/2$ new physics. The current $1 \sigma$ hint with large central value
 already favors  the $\rep{3}+\rep{15}$ Hamiltonian, an example of which is provided by  \cite{Hochberg:2011ru}. The operator enhancement 
 from the current fit 
 is, given the uncertainties, in the ballpark of what is required to explain the current anomalies in the top sector.
  \item[{\it vi}] We predict that $a_{CP}^{\mathrm dir}(D^0 \to K_S K_S)$ is enhanced with respect to $\Delta a_{CP}^{\rm dir}$, see Eq.~(\ref{eq:KsKs}).
 \end{itemize}

Future fits with improved data on charm CP violation \cite{Bediaga:2012py,Aushev:2010bq,O'Leary:2010af,Asner:2008nq} will shed light on the origin of CP and flavor violation in charm.


{\bf Note added: } While this work has been completed, a related study  
appeared \cite{Grossman:2012ry}.

\begin{acknowledgments}
This project is supported in part by the German-Israeli foundation for
scientific research and development (GIF) and 
the Bundesministerium f\"ur Bildung und Forschung (BMBF).
GH gratefully acknowledges the kind hospitality of the theory group at DESY Hamburg, where parts of this work have been done.
StS thanks Christian Hambrock for useful exchanges.
 
\end{acknowledgments}
 
\bigskip
 
\appendix

\section{Subtracting indirect CP asymmetries 
\label{app:removeMixing}}

Decays with a  $K_{S,L}$ in the final state receive an additional contribution to the CP asymmetry from kaon mixing. To isolate the direct
CP asymmetry, we subtract the mixing contribution $\propto\delta_L=\frac{2 Re \varepsilon}{1+|\varepsilon|^2}$.
For a decay with a $K^0$ ($\bar{K}^0$) in the flavor final state, the mixing contribution to 
its  CP asymmetry is given by $A_{CP}^{K^0} = \delta_L$ ($A_{CP}^{\bar{K}^0} = -\delta_L$)
\cite{Cenci:2012uy}.
Therefore, 
\begin{align}
a^{\mathrm dir}_{CP}(D^+ \rightarrow K_S K^+)  &= A_{CP}(D^+ \rightarrow K_S K^+)  + \delta_L   ,\\
a^{\mathrm dir}_{CP}(D_s\rightarrow K_S \pi^+) &= A_{CP}(D_s\rightarrow K_S \pi^+) - \delta_L\,  .
\end{align}
As pointed out in \cite{Grossman:2011zk}, the actual influence of kaon mixing 
depends on the experiment, due to its dependence on the kaon decay time. 
In the most recent analyses  \cite{Cenci:2012uy,CKM12:Ko} this effect  is accounted for.

To obtain $a^{\mathrm{dir}}_{CP}(D^0\rightarrow \pi^0 \pi^0)$, we took into account the effect of $D^0-\bar{D}^0$ mixing in the same way as kaon mixing
and subtracted $a_{CP}^{\mathrm{ind}}$.  

Indirect CP violation in both kaon and charm mixing is discarded in 
$A_{CP}(D^0 \rightarrow K_S K_S)$, as for this mode the experimental uncertainties are much larger than these effects. 

To calculate $\Sigma A_{CP}(K^+K^-,\pi^+\pi^-)$, we average the data from CDF and the B-factories, taking into account the correlations between $A_{CP}(D^0\rightarrow K^+K^-)$ and $A_{CP}(D^0\rightarrow\pi^+\pi^-)$.  
In order to obtain $\Sigma a_{CP}^{\mathrm{dir}}(K^+K^-,\pi^+\pi^-)$, we subtract the contribution from indirect CP violation. 
We find the correlation coefficient between $\Delta  a_{CP}^{\mathrm{dir}}$ and $\Sigma a_{CP}^{\mathrm{dir}}$ to be only $\sim 5\%$, which we can safely neglect in our analysis. The reason for the small correlation lies in the small impact of the $B$-factory results and $a_{CP}^{\mathrm{ind}}$ on the average of $\Delta a_{CP}^{\mathrm{dir}}(K^+K^-,\pi^+\pi^-)$.

\section{Amplitudes including $\mathbf{SU(3)}$ breaking}
In this appendix, the coefficient tables for the $SU(3)$-breaking parts of the amplitudes are given. In Table~\ref{tab:SU3breakoctetpre} we list the coefficients for the full set of matrix elements, before reducing the basis to include only physical ones. The coefficients for the latter are given in Table~\ref{tab:SU3breakoctet}.
\begin{table*}[hbt]
\begin{scriptsize}
\begin{center}
\begin{tabular}{l|c|c|c|c|c|c|c|c|c|c|c|c|c|c|c}
\hline \hline
Decay $d$&  $B_1^{3_1}$ & $B_1^{3_2}$  &  $B_8^{3_1}$ & $B_8^{3_2}$   & $B_8^{\bar{6}_{1}}$   & $B_8^{\bar{6}_{2}}$  & $B_8^{15_{1}}$  & $B_8^{15_{2}}$  &  $B_8^{15_{3}}$ & $B_{27}^{15_{1}}$  & $B_{27}^{15_{2}}$  & $B_{27}^{15_{3}}$  & $B_{27}^{24_{1}}$  & $B_{27}^{24_{2}}$  &  $B_{27}^{42}$  \\\hline\hline
\multicolumn{16}{c}{SCS} \\\hline\hline
$D^0\rightarrow K^+ K^-$  & $\frac{1}{4 \sqrt{10}}$ & $\frac{1}{8}$ & $\frac{1}{10 \sqrt{2}}$ & $\frac{1}{4 \sqrt{5}}$ & $\frac{1}{10}$ & $-\frac{1}{10 \sqrt{2}}$ & $-\frac{7}{10 \sqrt{122}}$ & $\frac{\sqrt{\frac{3}{122}}}{5}$ & $-\frac{1}{20}$ & $-\frac{31}{20 \sqrt{122}}$ & $-\frac{17}{20 \sqrt{366}}$ & $\frac{7}{40}$ & $-\frac{1}{10 \sqrt{6}}$ & $\frac{1}{10 \sqrt{2}}$ & $-\frac{13}{20 \sqrt{42}}$    \\\hline  
$D^0\rightarrow \pi^+ \pi^- $ & $\frac{1}{4 \sqrt{10}}$ & $\frac{1}{8}$ & $\frac{1}{10 \sqrt{2}}$ & $\frac{1}{4 \sqrt{5}}$ & $-\frac{1}{10}$ & $\frac{1}{10 \sqrt{2}}$ & $-\frac{11}{10 \sqrt{122}}$ & $-\frac{2 \sqrt{\frac{2}{183}}}{5}$ & $\frac{3}{20}$ & $-\frac{23}{20 \sqrt{122}}$ & $\frac{11}{20 \sqrt{366}}$ & $-\frac{1}{40}$ & $\frac{1}{10 \sqrt{6}}$ & $-\frac{1}{10 \sqrt{2}}$ & $\frac{\sqrt{\frac{7}{6}}}{20}$  \\\hline  
$D^0\rightarrow \bar{K}^0 K^0$ & $-\frac{1}{4 \sqrt{10}}$ & $-\frac{1}{8}$ & $\frac{1}{5 \sqrt{2}}$ & $\frac{1}{2 \sqrt{5}}$ & $0$ & $0$ & $-\frac{9}{5 \sqrt{122}}$ & $-\frac{1}{5 \sqrt{366}}$ & $\frac{1}{10}$ & $-\frac{9}{20 \sqrt{122}}$ & $-\frac{1}{20 \sqrt{366}}$ & $\frac{1}{40}$ & $-\frac{1}{2 \sqrt{6}}$ & $-\frac{1}{2 \sqrt{2}}$ & $\frac{19}{20 \sqrt{42}}$  \\\hline  
$D^0 \rightarrow \pi^0 \pi^0$ & $-\frac{1}{8 \sqrt{5}}$ & $-\frac{1}{8 \sqrt{2}}$ & $-\frac{1}{20}$ & $-\frac{1}{4 \sqrt{10}}$ & $\frac{1}{10 \sqrt{2}}$ & $-\frac{1}{20}$ & $\frac{11}{20 \sqrt{61}}$ & $\frac{2}{5 \sqrt{183}}$ & $-\frac{3}{20 \sqrt{2}}$ & $-\frac{57}{40 \sqrt{61}}$ & $\frac{7}{20 \sqrt{183}}$ & $\frac{1}{40 \sqrt{2}}$ & $\frac{1}{5 \sqrt{3}}$ & $\frac{1}{20}$ & $-\frac{1}{20 \sqrt{21}}$  \\\hline  
$D^+ \rightarrow \pi^0 \pi^+$ & $0$ & $0$ & $0$ & $0$ & $0$ & $0$ & $0$ & $0$ & $0$ & $-\frac{2 \left(1-\tilde\Delta\right)}{\sqrt{61}}$ & $\frac{5 \left(1-\tilde\Delta\right)  }{8 \sqrt{183}}$ & $0$ & $\frac{ 1-\tilde\Delta  }{4 \sqrt{3}}$ & $0$ & $\frac{ 1-\tilde\Delta }{8 \sqrt{21}}$    \\\hline
$D^+ \rightarrow \bar{K}^0 K^+$ & $0$ & $0$ & $\frac{3}{10 \sqrt{2}}$ & $\frac{3}{4 \sqrt{5}}$ & $\frac{1}{10}$ & $-\frac{1}{10 \sqrt{2}}$ & $\frac{7}{10 \sqrt{122}}$ & $-\frac{\sqrt{\frac{3}{122}}}{5}$ & $\frac{1}{20}$ & $-\frac{3 \sqrt{\frac{2}{61}}}{5}$ & $-\frac{23}{20 \sqrt{366}}$ & $\frac{1}{5}$ & $-\frac{1}{10 \sqrt{6}}$ & $-\frac{\sqrt{2}}{5}$ & $-\frac{19}{20 \sqrt{42}}$   \\\hline  
$D_s \rightarrow  K^0 \pi^+$ & $0$ & $0$ & $\frac{3}{10 \sqrt{2}}$ & $\frac{3}{4 \sqrt{5}}$ & $-\frac{1}{10}$ & $\frac{1}{10 \sqrt{2}}$ & $\frac{11}{10 \sqrt{122}}$ & $\frac{2 \sqrt{\frac{2}{183}}}{5}$ & $-\frac{3}{20}$ & $-\frac{3}{5 \sqrt{122}}$ & $\frac{19}{20 \sqrt{366}}$ & $-\frac{1}{10}$ & $-\frac{\sqrt{\frac{2}{3}}}{5}$ & $-\frac{1}{10 \sqrt{2}}$ & $-\frac{19}{20 \sqrt{42}}$  \\\hline  
$D_s \rightarrow  K^+ \pi^0$ & $0$ & $0$ & $-\frac{3}{20}$ & $-\frac{3}{4 \sqrt{10}}$ & $\frac{1}{10 \sqrt{2}}$ & $-\frac{1}{20}$ & $-\frac{11}{20 \sqrt{61}}$ & $-\frac{2}{5 \sqrt{183}}$ & $\frac{3}{20 \sqrt{2}}$ & $-\frac{17}{10 \sqrt{61}}$ & $\frac{\sqrt{\frac{3}{61}}}{20}$ & $\frac{1}{10 \sqrt{2}}$ & $-\frac{\sqrt{3}}{10}$ & $\frac{1}{20}$ & $-\frac{\sqrt{\frac{3}{7}}}{20}$   \\\hline\hline
\multicolumn{16}{c}{CF} \\\hline\hline
$D^0\rightarrow K^- \pi^+$ & $0$ & $0$ & $0$ & $0$ & $\frac{1}{5}$ & $\frac{1}{5 \sqrt{2}}$ & $-\frac{\sqrt{\frac{2}{61}}}{5}$ & $-\frac{7}{5 \sqrt{366}}$ & $-\frac{1}{5}$ & $\frac{\sqrt{\frac{2}{61}}}{5}$ & $\frac{7}{5 \sqrt{366}}$ & $\frac{1}{5}$ & $\frac{1}{20 \sqrt{6}}$ & $\frac{1}{20 \sqrt{2}}$ & $-\frac{1}{2 \sqrt{42}}$  \\\hline  
$D^0\rightarrow \bar{K}^0 \pi^0$  &  $0$ & $0$ & $0$ & $0$ & $-\frac{1}{5 \sqrt{2}}$ & $-\frac{1}{10}$ & $\frac{1}{5 \sqrt{61}}$ & $\frac{7}{10 \sqrt{183}}$ & $\frac{1}{5 \sqrt{2}}$ & $\frac{3}{10 \sqrt{61}}$ & $\frac{7 \sqrt{\frac{3}{61}}}{20}$ & $\frac{3}{10 \sqrt{2}}$ & $-\frac{\sqrt{3}}{20}$ & $-\frac{3}{20}$ & $0$    \\\hline  
$D^+ \rightarrow \bar{K}^0 \pi^+$ &  $0$ & $0$ & $0$ & $0$ & $0$ & $0$ & $0$ & $0$ & $0$ & $\frac{1}{\sqrt{122}}$ & $\frac{7}{2 \sqrt{366}}$ & $\frac{1}{2}$ & $-\frac{1}{4 \sqrt{6}}$ & $-\frac{1}{4 \sqrt{2}}$ & $-\frac{1}{2 \sqrt{42}}$  \\\hline
$D_s \rightarrow \bar{K}^0 K^+$ &  $0$ & $0$ & $0$ & $0$ & $-\frac{1}{5}$ & $-\frac{1}{5 \sqrt{2}}$ & $-\frac{\sqrt{\frac{2}{61}}}{5}$ & $-\frac{7}{5 \sqrt{366}}$ & $-\frac{1}{5}$ & $\frac{\sqrt{\frac{2}{61}}}{5}$ & $\frac{7}{5 \sqrt{366}}$ & $\frac{1}{5}$ & $\frac{1}{5 \sqrt{6}}$ & $\frac{1}{5 \sqrt{2}}$ & $\frac{1}{\sqrt{42}}$   \\\hline\hline
\multicolumn{16}{c}{DCS} \\\hline\hline
$D^0 \rightarrow  K^+ \pi^-$ &  $0$ & $0$ & $0$ & $0$ & $0$ & $-\frac{\sqrt{2}}{5}$ & $\frac{2 \sqrt{\frac{2}{61}}}{5}$ & $\frac{7 \sqrt{\frac{2}{183}}}{5}$ & $0$ & $-\frac{2 \sqrt{\frac{2}{61}}}{5}$ & $-\frac{7 \sqrt{\frac{2}{183}}}{5}$ & $0$ & $-\frac{1}{4 \sqrt{6}}$ & $\frac{3}{20 \sqrt{2}}$ & $-\frac{1}{2 \sqrt{42}}$  \\\hline  
$D^0 \rightarrow K^0 \pi^0$ & $0$ & $0$ & $0$ & $0$ & $0$ & $\frac{1}{5}$ & $-\frac{2}{5 \sqrt{61}}$ & $-\frac{7}{5 \sqrt{183}}$ & $0$ & $-\frac{3}{5 \sqrt{61}}$ & $-\frac{7 \sqrt{\frac{3}{61}}}{10}$ & $0$ & $-\frac{\sqrt{3}}{8}$ & $-\frac{3}{40}$ & $0$   \\\hline
$D^+ \rightarrow K^0\pi^+ $ & $0$ & $0$ & $0$ & $0$ & $0$ & $\frac{\sqrt{2}}{5}$ & $\frac{2 \sqrt{\frac{2}{61}}}{5}$ & $\frac{7 \sqrt{\frac{2}{183}}}{5}$ & $0$ & $-\frac{2 \sqrt{\frac{2}{61}}}{5}$ & $-\frac{7 \sqrt{\frac{2}{183}}}{5}$ & $0$ & $-\frac{1}{4 \sqrt{6}}$ & $-\frac{3}{20 \sqrt{2}}$ & $-\frac{1}{2 \sqrt{42}}$   \\\hline
$D^+ \rightarrow K^+ \pi^0$ & $0$ & $0$ & $0$ & $0$ & $0$ & $-\frac{1}{5}$ & $-\frac{2}{5 \sqrt{61}}$ & $-\frac{7}{5 \sqrt{183}}$ & $0$ & $-\frac{3}{5 \sqrt{61}}$ & $-\frac{7 \sqrt{\frac{3}{61}}}{10}$ & $0$ & $-\frac{\sqrt{3}}{8}$ & $\frac{3}{40}$ & $0$   \\\hline
$D_s \rightarrow K^0 K^+ $ & $0$ & $0$ & $0$ & $0$ & $0$ & $0$ & $0$ & $0$ & $0$ & $-\sqrt{\frac{2}{61}}$ & $-\frac{7}{\sqrt{366}}$ & $0$ & $\frac{1}{2 \sqrt{6}}$ & $0$ & $\frac{1}{\sqrt{42}}$    \\\hline \hline
\end{tabular}
\caption{The coefficients $c_{d;ij}$ of the $SU(3)$-breaking decomposition given in Eqs.~(\ref{eq:su3breakamplitudes})-(\ref{eq:su3breakamplitudes3}) without reparametrizations.  
\label{tab:SU3breakoctetpre}}
\end{center}
\end{scriptsize}
\end{table*}

\begin{table*}
\begin{center}
\begin{tabular}{l|c|c|c|c|c|c|c|c}
\hline \hline
Decay $d$&  $B_1^{3}$ & $B_8^{3}$    & $B_8^{\bar{6}_{1}}$   & $B_8^{15_{1}}$  & $B_8^{15_{2}}$  &  $B_{27}^{15_{1}}$  & $B_{27}^{15_{2}}$  & $B_{27}^{24_{1}}$ \\\hline\hline
\multicolumn{9}{c}{SCS} \\\hline\hline
$D^0\rightarrow K^+ K^-$  & $\frac{\sqrt{\frac{421}{35}}}{16}$ & $\frac{\sqrt{\frac{3937}{7}}}{160}$ & $\frac{\sqrt{\frac{2869}{7}}}{80}$ & $-\frac{\sqrt{9316783}}{29280}$ & $\frac{\sqrt{\frac{2613}{2}}}{610}$ & $-\frac{31 \sqrt{\frac{5281}{7}}}{4880}$ & $-\frac{17 \sqrt{\frac{151}{21}}}{610}$ & $-\frac{1}{5 \sqrt{21}}$  \\\hline 
$D^0\rightarrow \pi^+ \pi^- $ & $\frac{\sqrt{\frac{421}{35}}}{16}$ & $\frac{\sqrt{\frac{3937}{7}}}{160}$ & $-\frac{\sqrt{\frac{2869}{7}}}{80}$ & $-\frac{11 \sqrt{\frac{1330969}{7}}}{29280}$ & $-\frac{\sqrt{\frac{1742}{3}}}{305}$ & $-\frac{23 \sqrt{\frac{5281}{7}}}{4880}$ & $\frac{11 \sqrt{\frac{151}{21}}}{610}$ & $\frac{1}{5 \sqrt{21}}$ \\\hline
$D^0\rightarrow \bar{K}^0 K^0$ & $-\frac{\sqrt{\frac{421}{35}}}{16}$ & $\frac{\sqrt{\frac{3937}{7}}}{80}$ & $0$ & $-\frac{3 \sqrt{\frac{1330969}{7}}}{4880}$ & $-\frac{\sqrt{\frac{871}{6}}}{610}$ & $-\frac{9 \sqrt{\frac{5281}{7}}}{4880}$ & $-\frac{\sqrt{\frac{151}{21}}}{610}$ & $-\frac{1}{\sqrt{21}}$ \\\hline
$D^0 \rightarrow \pi^0 \pi^0$ & $-\frac{\sqrt{\frac{421}{70}}}{16}$ & $-\frac{\sqrt{\frac{3937}{14}}}{160}$ & $\frac{\sqrt{\frac{2869}{14}}}{80}$ & $\frac{11 \sqrt{\frac{1330969}{14}}}{29280}$ & $\frac{\sqrt{\frac{871}{3}}}{305}$ & $-\frac{57 \sqrt{\frac{5281}{14}}}{4880}$ & $\frac{\sqrt{\frac{1057}{6}}}{305}$ & $\frac{2 \sqrt{\frac{2}{21}}}{5}$ \\\hline
$D^+ \rightarrow \pi^0 \pi^+$ & $0$ & $0$ & $0$ & $0$ & $0$ & $-\frac{\sqrt{\frac{5281}{14}} ( 1 - \tilde\Delta  )}{61  }$ & $\frac{5 \sqrt{\frac{151}{42}} (1 - \tilde\Delta )}{122  }$ & $\frac{1  -\tilde\Delta }{\sqrt{42} }$ \\\hline
$D^+ \rightarrow \bar{K}^0 K^+$ &  $0$ & $\frac{3 \sqrt{\frac{3937}{7}}}{160}$ & $\frac{\sqrt{\frac{2869}{7}}}{80}$ & $\frac{\sqrt{9316783}}{29280}$ & $-\frac{\sqrt{\frac{2613}{2}}}{610}$ & $-\frac{3 \sqrt{\frac{5281}{7}}}{610}$ & $-\frac{23 \sqrt{\frac{151}{21}}}{610}$ & $-\frac{1}{5 \sqrt{21}}$ \\\hline
$D_s \rightarrow  K^0 \pi^+$ & $0$ & $\frac{3 \sqrt{\frac{3937}{7}}}{160}$ & $-\frac{\sqrt{\frac{2869}{7}}}{80}$ & $\frac{11 \sqrt{\frac{1330969}{7}}}{29280}$ & $\frac{\sqrt{\frac{1742}{3}}}{305}$ & $-\frac{3 \sqrt{\frac{5281}{7}}}{1220}$ & $\frac{19 \sqrt{\frac{151}{21}}}{610}$ & $-\frac{4}{5 \sqrt{21}}$ \\\hline
$D_s \rightarrow  K^+ \pi^0$ & $0$ & $-\frac{3 \sqrt{\frac{3937}{14}}}{160}$ & $\frac{\sqrt{\frac{2869}{14}}}{80}$ & $-\frac{11 \sqrt{\frac{1330969}{14}}}{29280}$ & $-\frac{\sqrt{\frac{871}{3}}}{305}$ & $-\frac{17 \sqrt{\frac{5281}{14}}}{1220}$ & $\frac{\sqrt{\frac{453}{14}}}{305}$ & $-\frac{\sqrt{\frac{6}{7}}}{5}$ \\\hline
\multicolumn{9}{c}{CF} \\\hline\hline
$D^0\rightarrow K^- \pi^+$ & $0$ & $0$ & $\frac{\sqrt{\frac{2869}{7}}}{40}$ & $-\frac{\sqrt{\frac{1330969}{7}}}{7320}$ & $-\frac{7 \sqrt{\frac{871}{6}}}{610}$ & $\frac{\sqrt{\frac{5281}{7}}}{610}$ & $\frac{2 \sqrt{\frac{1057}{3}}}{305}$ & $\frac{1}{10 \sqrt{21}}$ \\\hline
$D^0\rightarrow \bar{K}^0 \pi^0$  & $0$ & $0$ & $-\frac{\sqrt{\frac{2869}{14}}}{40}$ & $\frac{\sqrt{\frac{1330969}{14}}}{7320}$ & $\frac{7 \sqrt{\frac{871}{3}}}{1220}$ & $\frac{3 \sqrt{\frac{5281}{14}}}{1220}$ & $\frac{\sqrt{\frac{3171}{2}}}{305}$ & $-\frac{\sqrt{\frac{3}{14}}}{5}$ \\\hline
$D^+ \rightarrow \bar{K}^0 \pi^+$ & $0$ & $0$ & $0$ & $0$ & $0$ & $\frac{\sqrt{\frac{5281}{7}}}{244}$ & $\frac{\sqrt{\frac{1057}{3}}}{61}$ & $-\frac{1}{2 \sqrt{21}}$ \\\hline
$D_s \rightarrow \bar{K}^0 K^+$ &  $0$ & $0$ & $-\frac{\sqrt{\frac{2869}{7}}}{40}$ & $-\frac{\sqrt{\frac{1330969}{7}}}{7320}$ & $-\frac{7 \sqrt{\frac{871}{6}}}{610}$ & $\frac{\sqrt{\frac{5281}{7}}}{610}$ & $\frac{2 \sqrt{\frac{1057}{3}}}{305}$ & $\frac{2}{5 \sqrt{21}}$ \\\hline
\multicolumn{9}{c}{DCS} \\\hline\hline
$D^0 \rightarrow  K^+ \pi^-$ & $0$ & $0$ & $0$ & $\frac{\sqrt{\frac{1330969}{7}}}{3660}$ & $\frac{7 \sqrt{\frac{871}{6}}}{305}$ & $-\frac{\sqrt{\frac{5281}{7}}}{305}$ & $-\frac{4 \sqrt{\frac{1057}{3}}}{305}$ & $-\frac{1}{2 \sqrt{21}}$ \\\hline
$D^0 \rightarrow K^0 \pi^0$ & $0$ & $0$ & $0$ & $-\frac{\sqrt{\frac{1330969}{14}}}{3660}$ & $-\frac{7 \sqrt{\frac{871}{3}}}{610}$ & $-\frac{3 \sqrt{\frac{5281}{14}}}{610}$ & $-\frac{\sqrt{6342}}{305}$ & $-\frac{\sqrt{\frac{3}{14}}}{2}$ \\\hline
$D^+ \rightarrow K^0\pi^+ $ & $0$ & $0$ & $0$ & $\frac{\sqrt{\frac{1330969}{7}}}{3660}$ & $\frac{7 \sqrt{\frac{871}{6}}}{305}$ & $-\frac{\sqrt{\frac{5281}{7}}}{305}$ & $-\frac{4 \sqrt{\frac{1057}{3}}}{305}$ & $-\frac{1}{2 \sqrt{21}}$ \\\hline
$D^+ \rightarrow K^+ \pi^0$ & $0$ & $0$ & $0$ & $-\frac{\sqrt{\frac{1330969}{14}}}{3660}$ & $-\frac{7 \sqrt{\frac{871}{3}}}{610}$ & $-\frac{3 \sqrt{\frac{5281}{14}}}{610}$ & $-\frac{\sqrt{6342}}{305}$ & $-\frac{\sqrt{\frac{3}{14}}}{2}$ \\\hline
$D_s \rightarrow K^0 K^+ $ & $0$ & $0$ & $0$ & $0$ & $0$ & $-\frac{\sqrt{\frac{5281}{7}}}{122}$ & $-\frac{2 \sqrt{\frac{1057}{3}}}{61}$ & $\frac{1}{\sqrt{21}}$ \\\hline\hline 
\end{tabular}
\caption{The coefficients $c_{d;ij}$ of the physical $SU(3)$-breaking decomposition  used in our analysis, see
 Section \ref{sec:su3break}.}
\label{tab:SU3breakoctet}
\end{center}
\end{table*}

\section{\label{app:futdata} Future data scenario}
To obtain estimates for the experimental uncertainties in the future data scenario, we make the following assumptions:
we assume that the systematic uncertainty on $\Delta a_{CP}^{\rm dir}$ given in \cite{Aaij:2011in} improves by a factor $\sim 2$ and dominates the statistical uncertainty; furthermore, we assume several CP asymmetries to be measured with the same precision as $\Delta a_{CP}^{\mathrm{dir}}$, see Table~\ref{tab:future}. 
The prospect for the uncertainty of the strong phase is taken from \cite{CKM12:Asner}. 

As in the $\Delta U=1$ model there is NP in the $\bar{s} s$ coupling, we assume
that $a_{CP}^{\mathrm{dir}}(D^0\rightarrow K^+ K^-)$ and $a_{CP}^{\mathrm{dir}}(D^+\rightarrow K_S K^+)$
are enhanced, whereas $a_{CP}^{\mathrm{dir}}(D^0\rightarrow \pi^+\pi^-)$ and 
$a_{CP}^{\mathrm{dir}}(D_s\rightarrow K_S \pi^+) \sim 0$. 
Furthermore, in the triplet model the observables $a_{CP}^{\mathrm{dir}}(D^+\rightarrow K_S K^+)$,
$a_{CP}^{\mathrm{dir}}(D_s\rightarrow K_S \pi^+)$ and $a_{CP}^{\mathrm{dir}}(D_s\rightarrow K^+\pi^0)$
are all determined uniquely by $A_8^3$, whereas $A_1^3$ does not contribute here. 
It is therefore difficult in the triplet model to account for  these three CP asymmetries not being of the same
order of magnitude. Consequently, we  assume $a_{CP}^{\mathrm{dir}}(D_s\rightarrow K^+\pi^0)$ to stay large.
All assumed central values are within
the $2\sigma$ interval of the current measurements listed in Table~\ref{tab:data}.

Note that the expectation that the operator $\bar s c \bar u s$ in the $\Delta U=1$ model contributes mainly to matrix elements with kaons is not a statement from $SU(3)$. It would amount to additional dynamical input, which we avoid in this work. Such assumptions could be implemented by assuming relations between matrix elements.

\newpage

\end{document}